\documentclass[aps,pre,preprint,groupedaddress]{revtex4-2}

\usepackage{graphicx}
\usepackage{amsmath}
\usepackage{amssymb}
\usepackage{booktabs}

\begin{document}


\title{Convectons in unbalanced natural doubly diffusive convection}


\author{J. Tumelty}

\author{C. Beaume}
\email[]{c.m.l.beaume@leeds.ac.uk}

\author{A. M. Rucklidge}
\email[]{a.m.rucklidge@leeds.ac.uk}
\affiliation{School of Mathematics, University of Leeds, Leeds LS2 9JT (UK)}


\date{\today}

\begin{abstract}
Fluids subject to both thermal and compositional variations can undergo doubly diffusive convection when these properties both affect the fluid density and diffuse at different rates. 
In natural doubly diffusive convection, the gradients of temperature and salinity are aligned with each other and orthogonal to gravity.
The resulting buoyancy-driven flows are known to lead to the formation of a variety of patterns, including spatially localized states of convection surrounded by quiescent fluid, which are known as convectons.
Localized pattern formation in natural doubly diffusive convection has been studied under a specific balance where the effects of temperature and salinity changes are opposite but of equal intensity on the fluid density.
In this case, a steady conduction state exists and convectons bifurcate from it.
The aforementioned buoyancy balance underpins our knowledge of this pattern formation but it is an ideal case that can hardly be met experimentally or in nature. 
This article addresses how localized pattern formation in natural doubly diffusive convection is affected by departures from the balanced case.
In particular, the absence of a conduction state leads to the unfolding of the bifurcations to convectons.
In thermally dominated regimes, the background flow promotes localized states with convection rolls attached to the end-walls, known as anticonvectons, and the existence of these states is found to be related to the emergence of convectons.
The results presented here shed light on the convecton robustness against changes in the buoyancy ratio and extend the scope of our understanding of localized pattern formation in fluid systems.
\end{abstract}



\maketitle

\section{Introduction}
Thermal convection provides a great framework for the study of pattern formation in fluid systems.
Rayleigh--B\'enard convection, where a horizontal fluid layer is heated from below, is widely used as an example of convection roll formation \cite{chandrasekhar81,koschmieder93} but it is also extensively studied for more complex phenomena like the formation of a variety of patterns on the horizontal plane \cite{Golubitsky84} and transitions in a cubic domain \cite{pallares01,puigjaner06} or in cylindrical geometries \cite{boronska10a,boronska10b}.
However, since Rayleigh--B\'enard convection is a supercritical problem, it has limited scope for multi-stability near onset.
In doubly diffusive convection, by contrast, the presence of a competition between the thermal and the solutal effects on the fluid density may change the system criticality and lead to rich dynamics close to onset \cite{alonso07}.
When the driving gradients of temperature and salinity are parallel to gravity, two important instabilities may arise and have been extensively studied by the astrophysical and geophysical fluid dynamics community: salt fingering \cite{stern69} and diffusive layering \cite{kelley03}.
These instabilities are not the only interesting phenomenological points of interest provided by doubly diffusive convection and other instabilities may be triggered when the buoyancy gradients are inclined, for example, in the vicinity of icebergs \cite{huppert81}.

We are interested here in the case of a vertical column of fluid where the gradients of temperature and salinity are aligned with each other and orthogonal to gravity.
Pattern formation in the resulting natural doubly diffusive convection system has been mostly studied in the subcritical case by looking at linear instabilities from the conduction state, for which the fluid is motionless \cite{ghorayeb1997double,xin1998bifurcation,beaume22}. 
Pioneering simulations in three dimensions revealed a complex bifurcation scenario \cite{bergeon02}.
Further, the discovery of multiple coexisting solutions and, in particular, of a steady state consisting in a spatially localized convection roll in the subcritical regime by Ghorayeb \& Mojtabi \cite{ghorayeb1997double} fostered interest in spatially extended domains.
The systematic understanding of localized states into a bifurcation scenario called {\it homoclinic snaking} \cite{burke2006localized}, a term first coined by Woods and Champneys \cite{woods99}, subsequently guided investigations and led to their first characterization in large, two-dimensional domains of doubly diffusive convection \cite{bergeon2008spatially}, where they are called {\it convectons}.
Further work focused on three-dimensional domains and identified a new instability responsible for secondary snaking nested within the better-known, primary snaking \cite{beaume2013convectons, beaume2018three}, localization in the transverse direction \cite{beaume2013nonsnaking} and a sudden route to chaos \cite{beaume20}.

In parallel, convectons have been studied in the doubly diffusive convection taking place in horizontal fluid layers driven by a vertical gradient of temperature and by the Soret effect, the solutal response to an applied thermal gradient.
In this system, usually referred to as {\it binary fluid convection}, Mercader {\it et al.} first computed the convecton bifurcation diagram and revealed that, in the presence of end-walls (using so-called ``closed container boundary conditions''), homoclinic snaking could be observed without bistability between the trivial state and a spatially periodic state \cite{mercader09}.
They subsequently identified new families of localized states: anticonvectons, which display a void in the center of the domain surrounded by convection rolls, and multiconvectons, which are bound states of convectons and anticonvectons \cite{mercader2011convectons}.
Other, more complex types of convectons have also been found, such as the periodic traveling pulse, which consists in a slow-moving envelope inside which convection rolls travel fast and outside which the fluid is motionless \cite{watanabe10,watanabe12}.
Convectons have also been found in magnetohydrodynamic convection \cite{blanchflower99}, Marangoni convection \cite{Assemat08}, binary fluid convection in a porous medium \cite{lojacono10} and rotating convection \cite{beaume13rotating}.

Besides their recent discovery in the supercritical regime \cite{tumelty2023toward}, doubly diffusive convectons have been studied in the subcritical regime where bistability is observed between a trivial conduction state and a domain-filling patterned state.
However, in the case of interest here, the existence of a conduction state is contingent on the presence of a physical balance: the fluid density needs to respond equally in amplitude but oppositely in sign to changes in its temperature and in its salinity.
This physical balance is controlled by a dimensionless parameter, the buoyancy ratio
\begin{equation}
N = -\dfrac{\rho_C  \Delta C}{\rho_T \Delta T},
\end{equation}
where $\rho_T$ ($\rho_C$) is the thermal (solutal) expansion coefficient and $\Delta T$ ($\Delta C$)  is the charateristic temperature (salinity) difference. 
The balanced case, $N=-1$, underpins all of our current understanding about convectons in doubly diffusive convection but it only represents an ideal case that is unlikely to be achieved experimentally or in nature.
The aim of our investigation is to understand how doubly diffusive convection is impacted by deviations from its balanced case, with particular focus on convectons and related localized states.
Knowledge of their robustness to parametric changes, particularly to those of~$N$, is critical to establish the relevance of a large body of literature.

We start in Section~\ref{sec:mathematical_formulation} by describing the mathematical formulation of the problem.
In Section~\ref{sec:N1}, we present the baseline results that are obtained in the balanced case ${N=-1}$.
In particular, we describe the primary bifurcations from the conduction state, as well as the bifurcation scenario of three families of convectons of interest: convectons, anticonvectons and multiconvectons.
Many of these states have never been computed in natural doubly diffusive convection and we have found that they play an important role in the emergence of convectons in the unbalanced case.
The main results of this paper are reported in Section~\ref{sec:Nneq1}, where we explain how the bifurcation scenario is impacted by the buoyancy ratio, identify how several types of localized states interact and explain the mechanism by which convectons are created in the thermally dominated regime ($N>-1$).
Finally, the paper ends with a discussion in Section~\ref{sec:discussion}.

\section{Mathematical formulation}\label{sec:mathematical_formulation}

\begin{figure}
  \centering
  \includegraphics[]{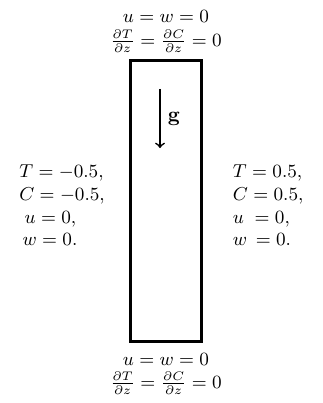}
  \caption[Representation of the non-dimensional closed cavity with no-slip boundary conditions]{Representation of the non-dimensional closed cavity with no-slip boundary conditions imposed at ${x = \pm0.5}$ and ${z = 0, L_z}$.}
  \label{fig:domain}
\end{figure}
In this paper, we consider the doubly diffusive convection of fluid confined within a closed rectangular cavity with thermal and solutal gradients imposed via the side-wall boundary conditions. 
The non-dimensionalized governing equations for this system read:
\begin{align}
  \dfrac{1}{Pr} \left(\dfrac{\partial \mathbf{u}}{\partial t} + \mathbf{u}\cdot \nabla \mathbf{u}\right) &= -\nabla p +\nabla^2 \mathbf{u} + Ra\left(T+NC\right)\mathbf{\hat{z}} , \label{eq:NS}\\
  \nabla \cdot \mathbf{u} &= 0, \label{eq:incomp}\\
  \dfrac{\partial T}{\partial t} + \mathbf{u}\cdot \nabla T &= \nabla^2 T, \label{eq:T}\\
  \dfrac{\partial C}{\partial t} + \mathbf{u}\cdot \nabla C &= \dfrac{1}{Le}\nabla^2 C, \label{eq:C}
\end{align}
where ${\bf u}$ is the fluid velocity, $p$ is its pressure, $T$ (resp. $C$) is the scaled temperature (resp. salinity) and ${\bf \hat{z}}$ is the upward-pointing unit vector.
The equations are parameterized by the Prandtl number $Pr$, the Lewis number $Le$, the buoyancy ratio $N$ and the Rayleigh number $Ra$.
In contrast to previous studies of convectons in similar configurations (see \cite{ bergeon2008spatially, beaume2011homoclinic, beaume2013convectons, tumelty2023toward}), we do not restrict our attention to the case where the buoyancy ratio satisfies $N=-1$.
Rather, we study the effect that a departure from this special value has on the system.
We fix ${Le = 5}$, ${Pr=1}$ and treat the Rayleigh number $Ra$ as a bifurcation parameter used to characterize steady states in the system.
A more physical description of the system and its non-dimensionalization can be found in the aforementioned papers.

The system (\ref{eq:NS})\textendash{}(\ref{eq:C}) is considered in the closed cavity that is bounded by the vertical side-walls at ${x = \pm 0.5}$ and horizontal end-walls at ${z = 0, L_z}$, as depicted in figure~\ref{fig:domain}.
We will consider ${L_z = 4\lambda_c}$, $5\lambda_c$ and $12\lambda_c$, where $\lambda_c \approx 2.48$ is the critical wavelength of the primary bifurcation with periodic boundary conditions in the $z$-direction \cite{xin1998bifurcation}.
The following boundary conditions are imposed on the cavity walls:
\begin{align}
  u = 0,&& w = 0, && -\frac{\partial p}{\partial x} + \frac{\partial^2 u}{\partial x^2}=0, && T = -0.5,& && C = -0.5&& \text{ on } & x = -0.5, \label{eq:bc0}\\
  u = 0,&& w = 0,&& -\frac{\partial p}{\partial x} + \frac{\partial^2 u}{\partial x^2}=0,&& T = \;\; 0.5,& && C = \;\; 0.5 && \text{ on }& x = 0.5, \label{eq:bc1}\\
  u = 0,&& w = 0,&& -\frac{\partial p}{\partial z} + \frac{\partial^2 u}{\partial z^2}=0,&& \frac{\partial T}{\partial z} =\;\; 0,& && \frac{\partial C}{\partial z} = \;\;0&& \text{ on }& z = 0, L_z, \label{eq:bcz0}
\end{align}
where, in addition to the fluid velocity vanishing on the four side-walls owing to the no-slip boundary conditions, the domain allows no thermal or solutal flux through the horizontal end-walls.

The results presented in this paper were obtained via numerical continuation using a spectral element numerical method based on a Gauss--Lobatto--Legendre discretization and supplemented by Stokes preconditioning (see \cite{mamun95,beaume2017adaptive} for further details).
Each element of the numerical domain was discretized using $25$ nodes in both the $x$- and $z$-directions, while the number of spectral elements $ne_z$ depends upon the vertical extent of the cavity (see table~\ref{tab:discretisation}). 
Validations were carried out using a higher resolution consisting of $30$ nodes in both directions.
The results are presented using bifurcation diagrams that, unless otherwise stated, present the total kinetic energy
\begin{equation}
  E = \frac{1}{2} \int_0^1\int_0^{L_z}\left( u^2+w^2\right)\, dz\,dx
\end{equation}
as a function of the Rayleigh number, or via the solution streamfunction.
These solution profiles use two scales to depict both weak and strong flows in the same figure: a logarithmic scale, showing black (clockwise flow) and grey (anticlockwise flow) streamlines; and a linear scale, showing red (clockwise flow) and blue (anticlockwise flow) streamlines.

\begin{table}
  \centering
  \begin{tabular}{cccc}
    \toprule
    $L_z$ & $ne_z$ & $n_{x}$ & $n_z$\\ 
    \midrule
    $\phantom{0}4\lambda_c\approx 9.93$  & $\phantom{0}8$ & $25$ & $25$ \\ 
    $\phantom{0}5\lambda_c\approx 12.4\phantom{0}$  &           $10$ & $25$ & $25$ \\ 
    $          12\lambda_c\approx 29.8\phantom{0}$  &           $24$ & $25$ & $25$ \\ 
    \bottomrule
  \end{tabular} 
  \caption[Discretizations used for numerical continuation]{Discretizations used for numerical continuation. The domain is discretized using $ne_z$ spectral elements, with each element containing $n_x$ nodes in the $x-$direction and $n_z$ nodes in the $z-$direction.}
  \label{tab:discretisation}
\end{table}

\section{Balanced system when $N = -1$}\label{sec:N1}

When $N=-1$, doubly diffusive convection admits a trivial state, called the conduction state, where the fluid is static ({\bf u} = {\bf 0}) and the temperature and salinity profiles are linear ($T=C=x$).
In this state, the buoyancy force (the last term in equation (\ref{eq:NS})) is zero since the thermal and solutal contributions are balanced, and the fluid density is uniform across the cavity.

The presence of end-walls at $z=0$ and $L_z$ creates fundamental differences between the system considered here and the more studied, spatially periodic one; in particular, this system is not equivariant with respect to continuous translations in the $z$-direction.
This precludes the emergence of pitchforks of revolution as an instability mechanism from the conduction state as found by Xin et al. \cite{xin1998bifurcation} in the spatially periodic case.
The presence of end-walls also excludes domain-filling, spatially periodic states from this system: the convection rolls that develop in doubly diffusive convection systems do not possess a horizontal section which is compatible with the boundary conditions we set at the end-walls.
For a more detailed discussion on the effect of boundary conditions on doubly diffusive convection pattern formation, we refer the reader to \cite{beaume2013convectons,beaume2013nonsnaking}.
These effects thus influence both where the branches of convectons bifurcate from and where they terminate, as will be seen in sections~\ref{sec:pri_s_N1} and \ref{sec:conv_N1}.
The following inversion symmetry about the center of the domain is, however, preserved:
\begin{equation}
S_{\Delta} : (x,z) \mapsto (-x,-z), \quad (u,w,T,C) \mapsto -(u,w,T,C).
\end{equation}
While this system admits states that break this inversion symmetry (for example, the two-dimensional equivalent to the two-roll wall-attached state found by Beaume et al. \cite{beaume2018three}), we largely only consider states that preserve this symmetry.

\subsection{Primary bifurcations of the conduction state}\label{sec:pri_s_N1}

We begin by detailing key properties of the balanced system ($N=-1$) with no-slip boundary conditions in $z$.
The conduction state can destabilise in either transcritical or pitchfork bifurcations \cite{ghorayeb1997double,xin1998bifurcation} when the buoyancy force driven by the side-wall heating overcomes viscous dissipation.
Transcritical bifurcations arise when $S_{\Delta}$-symmetric eigenmodes, which have a roll centered in the middle of the domain, become destabilizing, while pitchfork bifurcations arise when an eigenmode that breaks the $S_{\Delta}$-symmetry becomes destabilizing.
By varying the size of the domain, Ghorayeb and Mojtabi \cite{ghorayeb1997double} found that the ordering of the first set of transcritical and pitchfork bifurcations switches and that the two bifurcations occur in quicker succession in larger domains.
Thus, we start by illustrating these primary bifurcations in small domains (${L_z = 4\lambda_c\approx9.93}$ and ${L_z = 5\lambda_c\approx 12.4}$), before considering them in the larger domain with ${L_z = 12\lambda_c\approx 29.8}$, which will later be used to investigate the effects of taking $N\neq-1$ on spatially localized states.

\begin{figure}
  \centering
  \includegraphics[width=0.99\linewidth]{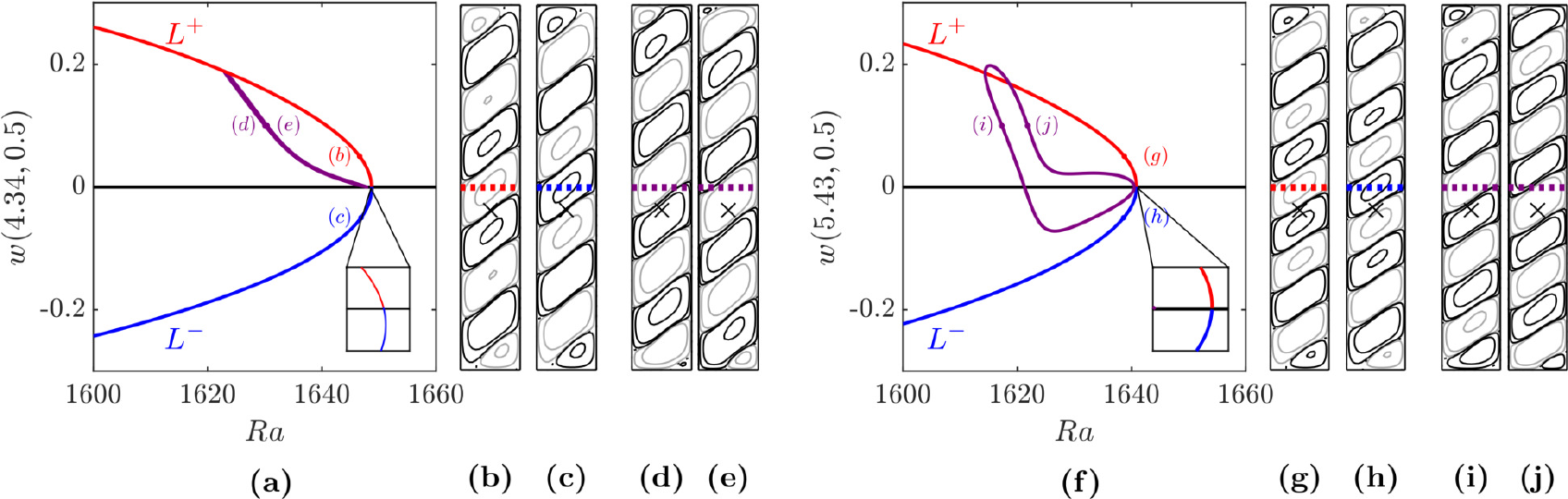}
  \caption[First primary pitchfork and transcritical bifurcations of the conduction state in domains with vertical extent ${L_z = 4\lambda_c}$ and ${L_z = 5\lambda_c}$]{First primary pitchfork and transcritical bifurcations of the conduction state in domains with vertical extent: (a)\textendash{}(e) ${L_z = 4\lambda_c}$ and (f)\textendash{}(j) ${L_z = 5\lambda_c}$.
    The bifurcation diagrams (a) and (f) show the vertical velocity of states at the point marked by the black crosses in the streamfunction plots (other panels) against the Rayleigh number.
    The following branches are shown: conduction state (black), $L^+$ (red), $L^-$ (blue) and branches of asymmetric states that bifurcate from the pitchfork bifurcation (purple).
    The insets provide a magnification of the first transcritical bifurcation.
    The remaining panels present the streamfunction of states marked in the bifurcation diagrams using contour values: $[-10^{-3}, -10^{-2}, -10^{-1}]$ (grey, anticlockwise); and $[10^{-3}, 10^{-2}, 10^{-1}]$ (black, clockwise). 
    Here, and in subsequent figures, the panels depicting the streamfunction are not to scale.
    The horizontal midline is shown using a red, blue or purple dotted line to indicate whether the state lies on the $L^+$~(blue), $L^+$~(red) or asymmetric (purple) branches.}
  \label{fig:N1_Lz4Lz5}
\end{figure}
\begin{figure}
  \centering
  \includegraphics[]{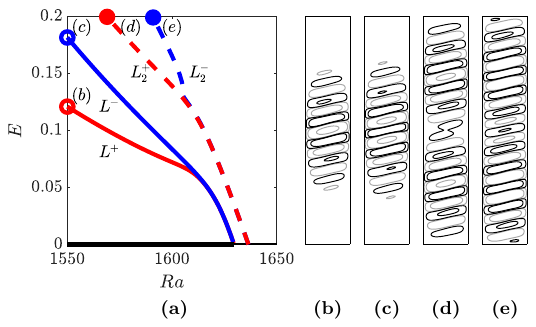}
  \caption[First two transcritical bifurcations for ${N = -1}$, ${Le = 5}$ and ${Pr = 1}$ in a domain with $L_z = 12\lambda_c$ and no-slip boundary conditions ]{First two transcritical bifurcations for ${N = -1}$, ${Le = 5}$ and ${Pr = 1}$ in a domain with ${L_z=12\lambda_c}$ and no-slip boundary conditions. 
    (a) Bifurcation diagram showing the branches that originate from these bifurcations. (b)\textendash{}(e) Streamfunctions along the four main branches: (b) $L^+$, (c) $L^-$, (d) $L_2^+$ and (e) $L_2^-$. 
    The contour values used were: $[-10^{-3}, -10^{-2}, -10^{-1}]$ (grey); and $[10^{-3}, 10^{-2}, 10^{-1}]$ (black).} 
  \label{fig:pribifN1}
\end{figure}

Figure~\ref{fig:N1_Lz4Lz5} presents the bifurcation diagrams showing the first two bifurcations of the conduction state when ${L_z = 4\lambda_c}$ (figure~\ref{fig:N1_Lz4Lz5}(a)\textendash{}(e)) and ${L_z=5\lambda_c}$ (figure~\ref{fig:N1_Lz4Lz5}(f)\textendash{}(j)).
In both of these domains, the conduction state first destabilizes in a pitchfork bifurcation and the resulting pair of branches consists of asymmetric states like those shown in figures~\ref{fig:N1_Lz4Lz5}(d), (e), (i) and (j).
These branches terminate on $L^+$ (red), the branch containing spatially modulated states with an anticlockwise central roll that bifurcates from the following transcritical bifurcation, from which $L^-$ (blue), the branch containing spatially modulated states with a clockwise central roll, also bifurcates.
The insets in figures~\ref{fig:N1_Lz4Lz5}(a) and (f) highlight the transcritical nature of this primary bifurcation and how the supercritical branch undergoes a saddle-node bifurcation shortly after onset, in an analogous manner to the three-dimensional behavior observed by Beaume et al. \cite{beaume2018three}.
The initially supercritical branch then heads towards lower Rayleigh numbers so that the behavior on a larger scale resembles that of a subcritical pitchfork bifurcation.
Which of the primary branches $L^+$ and $L^-$ is initially supercritical depends on the domain size
and we find that $L^-$ (blue) bifurcates supercritically when ${L_z = 4\lambda_c}$ while $L^+$ (red) bifurcates supercritically when ${L_z = 5\lambda_c}$.
The streamfunction panels (c) and (g) indicate that states on the supercritical branch possess clockwise outermost rolls, whereas the corresponding rolls are anticlockwise for the subcritical branch (see panels (b) and (h)), and this observation can likely be applied more generally to other domain sizes.

Figure~\ref{fig:pribifN1} presents the branches bifurcating from the first two transcritical bifurcations of the conduction state in the larger domain with ${L_z =12\lambda_c}$.
The conduction state first destabilizes in a pitchfork bifurcation (not shown owing to the short extent of the bifurcating branches) and, shortly after, in a transcritical bifurcation at ${Ra \approx 1629.6}$, where the branches $L^+$ and $L^-$ bifurcate subcritically and supercritically, respectively. 
These branches remain indistinguishable at small amplitude on the scale shown in figure~\ref{fig:pribifN1}(a) owing to the supercritical branch turning around at a saddle-node shortly after onset.

Due to the presence of no-slip boundary conditions on the horizontal end-walls, the critical eigenmode associated with the transcritical bifurcation is spatially modulated and vanishes at the ends of the domain.
This eigenmode possesses a central roll, which leads to the formation of an anticlockwise central roll for $L^+$ (figure~\ref{fig:pribifN1}(b)) and of a clockwise one for $L^-$ (figure~\ref{fig:pribifN1}(c)).
Owing to both the above amplitude modulation and to the nonlinear mechanism that favors anticlockwise convection rolls as the branches are followed towards lower Rayleigh numbers and larger amplitudes \cite{thangam82}, convectons develop with an odd number of localized rolls along $L^+$ and with an even number of localized rolls along $L^-$, as was previously found with periodic boundary conditions \cite{bergeon2008spatially}.

Beyond this first transcritical bifurcation, the conduction state undergoes further bifurcations associated with eigenmodes of different wavelengths. 
For example, the conduction state undergoes a further transcritical bifurcation at ${Ra \approx 1636.5}$, where branches of two-pulse states ($L_2^+$ and $L_2^-$) bifurcate. 
Like their single-pulse counterparts, states on these branches become amplitude modulated as each branch is followed towards larger amplitudes (see figures~\ref{fig:pribifN1}(d) and (e)).	
However, the primary bifurcation of the conduction state at ${Ra\approx1629.6}$ implies that the unstable eigenspace of these multi-pulse states is higher-dimensional than that of the $L^+$ or $L^-$ single-pulse states.

\subsection{Convectons}\label{sec:conv_N1}

\begin{figure}
  \centering
  \includegraphics[width=0.99\linewidth]{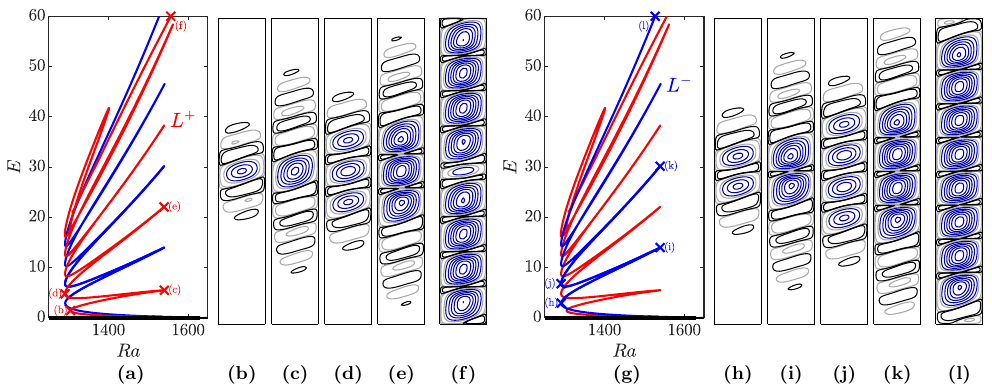}
  \caption[Convecton branches and streamfunction profiles for ${N=-1}$, ${Pr = 1}$ and ${Le = 5}$ in a domain with ${L_z = 12\lambda_c}$]{Convecton branches and streamfunction profiles for ${N=-1}$, ${Pr = 1}$ and ${Le = 5}$ in a domain with ${L_z = 12\lambda_c}$.
    (a) and (g) Bifurcation diagram showing branches corresponding to convectons with an odd ($L^+$) and even ($L^-$) number of strong anticlockwise central rolls, which are shown in red and blue, respectively.
    (b)\textendash{}(e) Streamfunction profiles of the first four saddle-nodes along $L^-$.
    (f) Streamfunction profile of large-amplitude state on $L^+$.
    (h)\textendash{}(k) Streamfunction profiles of the first four saddle-nodes along $L^+$.
    (l) Streamfunction profile of large-amplitude state on $L^-$.
    Contours are shown using linear (blue) and logarithmic (black and grey) scales and take the values: $[-10^{-3}, -10^{-2}, -10^{-1}]$ (grey); $[10^{-3}, 10^{-2}, 10^{-1}]$ (black); $[-0.2,-0.4,-0.6,...]$ (blue).}
  \label{fig:convecton_branch}
\end{figure}
In the larger domain, the convecton branches, $L^+$ and $L^-$, head away from the primary transcritical bifurcation of the conduction state and towards lower Rayleigh numbers before first turning around at a left saddle-node located at ${Ra \approx 1305}$ for $L^+$ (figure~\ref{fig:convecton_branch}(b)) or ${Ra \approx 1292}$ for $L^-$ (figure~\ref{fig:convecton_branch}(h)).
The amplitude modulation of the states increases as the branches approach their first left saddle-node, by which point convective motion has become spatially localized, as can be seen in figures~\ref{fig:convecton_branch}(b) and (h).
The central anticlockwise rolls also strengthen faster than the clockwise rolls, which maintain flow with a similar order of magnitude along these low-energy branch segments.

After the first left saddle-nodes, the two convecton branches proceed to undergo homoclinic snaking over the interval ${1291< Ra< 1540}$, as seen in figures~\ref{fig:convecton_branch}(a) and (g), which is similar to the pinning region when periodic boundary conditions are imposed in the vertical direction \cite{tumelty2023toward}.
The convecton rolls also exhibit similar strengthening and nucleation mechanisms as the branches are followed to large amplitude, as shown by the changes in the streamfunction plots between convectons at the first four saddle-nodes on $L^+$ (figures~\ref{fig:convecton_branch}(b)\textendash{}(e)) and $L^-$ (figures~\ref{fig:convecton_branch}(h)\textendash{}(k)).
Between the left and right saddle-nodes, the anticlockwise rolls strengthen and become larger. 
This reduces the distance between adjacent anticlockwise rolls as the interspersing weak clockwise rolls are squashed and may even split into two smaller rolls by the right saddle-node.
Outside of these central rolls, weak counterrotating rolls strengthen with increasing Rayleigh number while remaining modulated due to the boundary conditions.
As the branch further proceeds to the next left saddle-node, rolls decay with the exception of the anticlockwise rolls immediately adjacent to the large-amplitude central rolls, which strengthen.
As a result, by the next left saddle-node, the spatially localized structure comprises two more rolls.

The snaking behavior continues until anticlockwise rolls almost fill the domain.
Unlike with periodic boundary conditions, here, the convecton branches cannot terminate on branches of spatially periodic states but rather extend towards large amplitudes as defect states \cite{houghton2009homoclinic,bergeon2008eckhaus}.
On $L^+$, after reaching a nine-roll state at a left saddle-node (${Ra\approx 1302}$, ${E \approx 17.4}$), the central roll decays as the branch continues towards higher Rayleigh numbers while the remaining eight anticlockwise rolls strengthen, as seen in figure~\ref{fig:convecton_branch}(f).
The other convecton branch, $L^-$, exhibits alternative behavior, where an eight-roll state, like figure~\ref{fig:convecton_branch}(l), strengthens with increasing Rayleigh number until ${Ra \approx 2210}$, where each end anticlockwise roll splits into a pair of anticlockwise rolls and the branch returns to lower Rayleigh numbers as a ten-roll state (not shown).
The specific details of this large-amplitude behavior depends on the domain size, so will not be considered any further here, although we note that non-periodic boundary conditions are typically associated with more complicated behavior than periodic boundary conditions \cite{houghton2009homoclinic}. 

\subsection{Anticonvectons}\label{sec:antiN1}
Branches of anticonvectons\textemdash{}localized states where rolls are attached to both end-walls and separated by a void region\textemdash{}are a distinguishing feature of the system with no-slip boundary conditions.
Mercader et al. \cite{mercader2011convectons} found these states in binary fluid convection and attributed their presence to the absence of the translational invariance associated with periodic boundary conditions.
Branches of anticonvectons and convectons were found to undergo homoclinic snaking within the same pinning region. 
Anticonvectons also exist in natural doubly diffusive convection but, as will be seen, their properties tend to be more complex than those in binary fluid convection because of the dynamical difference between rolls with different sense of rotation.
\begin{figure}
  \centering
  \includegraphics[width=0.99\linewidth]{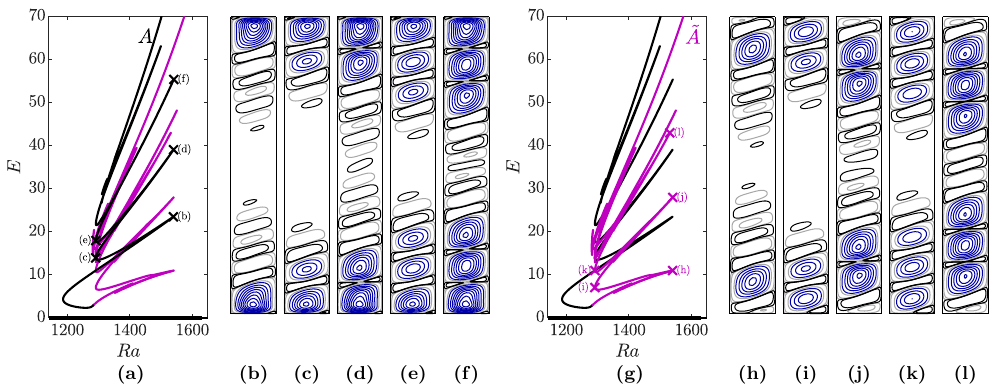}
  \caption[Anticonvecton branches and streamfunction profiles for ${N=-1}$, ${Pr = 1}$ and ${Le = 5}$]{Anticonvecton branches and streamfunction profiles for ${N=-1}$, ${Pr = 1}$ and ${Le = 5}$. 
    (a) and (g) Bifurcation diagrams corresponding to branches with two types of anticonvectons $A$ (black) and $\tilde{A}$ (purple) when ${N = -1}$.
    (b)\textendash{}(f) Streamfunctions of steady states at the saddle-nodes of the snaking anticonvecton branch $A$. (h)\textendash{}(l) Streamfunctions of steady states at the saddle-nodes of the snaking anticonvecton branch $\tilde{A}$.
    Contours are shown using linear (blue) and logarithmic (black and grey) scales and take the values: $[-10^{-3}, -10^{-2}, -10^{-1}]$ (grey); $[10^{-3}, 10^{-2}, 10^{-1}]$ (black); $[-0.2,-0.4,-0.6,...]$ (blue).}
  \label{fig:bdnsle5pr1n1_anticonvecton}
\end{figure}

Figure~\ref{fig:bdnsle5pr1n1_anticonvecton} shows two types of anticonvectons that lie on snaking branches that we refer to as $A$ (shown in black) and $\tilde{A}$ (shown in purple).
These branches appear to connect on the lowest branch segment, as they do when ${N\neq -1}$ (see figure~\ref{fig:trackanticonvectons}), but we could not confirm this here due to slow numerical convergence. 
Both anticonvecton branches proceed to snake upwards within the interval ${1291<Ra<1541}$, with both types of anticonvectons extending by a pair of anticlockwise rolls on the interior side of the existing convection pattern over a single snaking oscillation (e.g., compare panels (b), (d) and (f) or (h), (j) and (l)).
The oscillations continue until 6-roll states (shown in figures~\ref{fig:bdnsle5pr1n1_anticonvecton}(f) and (l)) are obtained and domain-size effects begin to be felt.

The form of the anticonvectons differ between the two branches. 
In states lying on the branch $A$ (figures~\ref{fig:bdnsle5pr1n1_anticonvecton}(b)\textendash{}(f)), the anticlockwise rolls attached to the end-walls are not elliptical like those in convectons, but rather appear to be squashed against the top or the bottom wall.
The end rolls weaken as the branch $A$ passes from the right saddle-nodes (e.g., (b), (d) and (f)) to the following left saddle-node (e.g., (c) and (e)). 
This pushes the anticlockwise rolls away from the top left and bottom right corners of the domain, thereby leaving space for the weak clockwise rolls in these corners to strengthen and grow.
The anticlockwise rolls that strengthen within the interior of the domain are not affected by the horizontal end-walls, however, and display similar forms to the rolls that nucleate in the convectons seen in figure~\ref{fig:convecton_branch}.

The $\tilde{A}$ anticonvectons also grow by nucleation of anticlockwise rolls on the interior side of the domain as the branch is continued from a right saddle-node to the next left one.
However, they also undergo more complicated changes.
The strong anticlockwise end rolls are no longer squashed against the horizontal walls, but rather separated from them by weaker rolls.
These weak rolls exhibit new behavior, referred to as a breathing mechanism, that is responsible for the vertical displacement of the large-amplitude convection rolls along the branch, as is shown in figures~\ref{fig:bdnsle5pr1n1_anticonvecton}(h)\textendash{}(l).
Following $\tilde{A}$ between a right saddle-node and the next left saddle-node, we see how the small, weak anticlockwise corner rolls weaken and disappear (e.g., between (h) and (i)).
The clockwise rolls between each weak corner roll and the adjacent strong anticlockwise roll also weaken as the Rayleigh number decreases to occupy only a small triangular region in the top left and bottom right corners of the domain at the left saddle-node. 
The weakening of this pair of rolls results in the strong anticlockwise rolls translating towards the end-walls as the branch is continued toward the left saddle-node.
This process occurs in reverse between each left saddle-node and the following right saddle-node, which leads to the weak end rolls strengthening and the strong anticlockwise rolls translating back towards the center of the domain.

\subsection{Multiconvectons}

\begin{figure}
  \centering
  \includegraphics[width=0.99\linewidth]{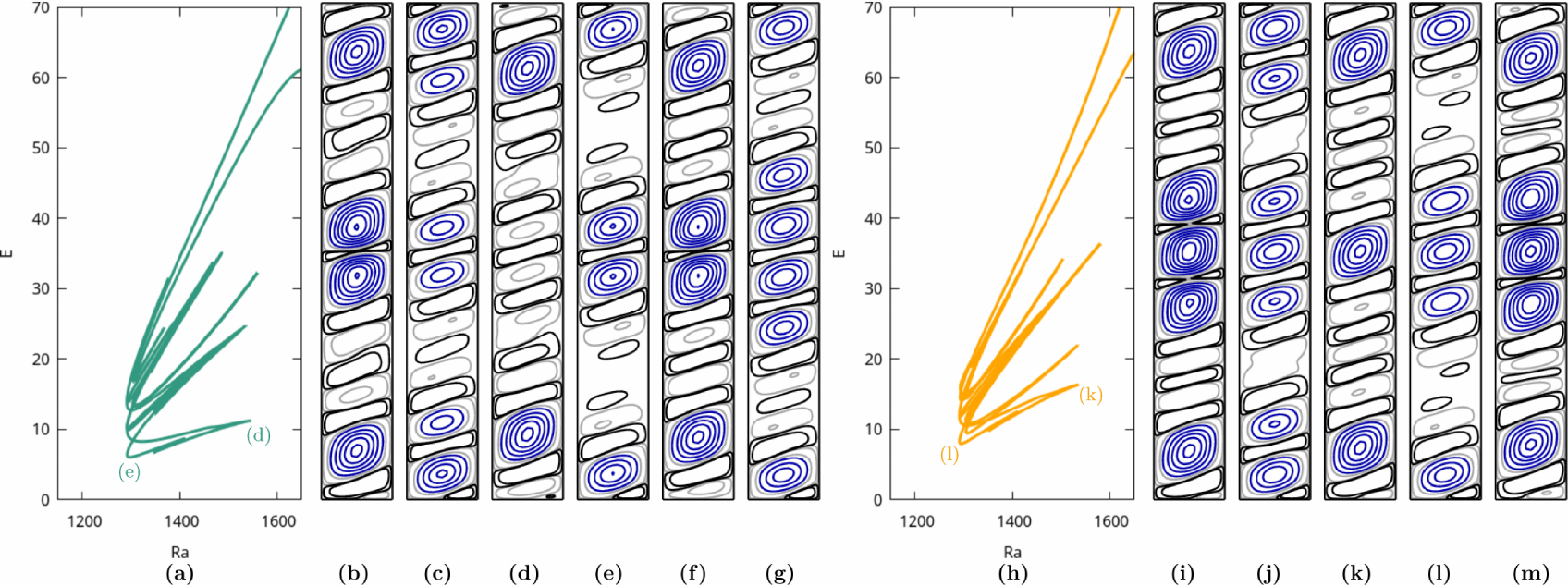}
  \caption{Multiconvecton branches and streamfunction profiles for $N=-1$, $Pr = 1$ and $Le = 5$. (a) and (h) Bifurcation diagrams for $\tilde{L}^-$ (green) and $\tilde{L}^+$ (orange). (b)--(g) Streamfunction of six consecutive outermost saddle-node steady states of the branch $\tilde{L}^-$, with (d) and (e) located on the lower segment of the branch. (i)--(m) Same as previously but for $\tilde{L}^+$. Only five steamfunction profiles are shown as the branch turns toward low energy beyond saddle-node state (m) and produces more complex behavior. Contours are shown using linear (blue) and logarithmic (black and grey) scales and take the values: $[-10^{-3}, -10^{-2}, -10^{-1}]$ (grey); $[10^{-3}, 10^{-2}, 10^{-1}]$ (black); $[-0.2,-0.4,-0.6,...]$ (blue).}
  \label{fig:hybridconvectons}
\end{figure}

Sufficiently large domains allow the existence of another family of steady states that combine convectons and anticonvectons.
These are characterized by convection rolls in the center of the domain and near its end-walls while the regions in between contain motionless fluid.
We refer to these states as {\it multiconvectons}.
Figure \ref{fig:hybridconvectons} shows the simplest multiconvectons that we found in doubly diffusive convection, as well as their bifurcation diagram.
To remain consistent with the rest of this paper, these are represented in a domain of size $12 \lambda_c$, leaving little room for the development of these heteroclinic connections.
As a result, what we mostly observe are states where the transitions from conduction to convection are seen via spatial modulation and are less obvious than they would be with sharply defined heteroclinic connections.
The branch shown in green, $\tilde{L}^-$, consists of multiconvectons with an even number of central rolls while the branch shown in orange, $\tilde{L}^+$, consists of multiconvectons with an odd number of central rolls.
Each of these branches has a similar snaking bifurcation diagram to the one produced by $A$ and $\tilde{A}$ except that the bottom left saddle-node of both branches is aligned with the left boundary of the pinning region and, due to the fact that multiconvectons initially contain more convection rolls than the less energetic anticonvectons or convectons, multiconvecton branches are found at larger values of the kinetic energy.

The multiconvecton at the lowest energy point on $\tilde{L}^-$ (figure \ref{fig:hybridconvectons}(e)) combines an $\tilde{A}$ ``anticonvecton'' with a single roll near each of the end-walls and a two-roll ``convecton'' at the center of the domain.
The multiconvectons on this branch are found to exhibit growth mechanisms observed in both the convecton branch $L^-$ and the anticonvecton branch $\tilde{A}$.
For example, following $\tilde{L}^-$ over a single snaking oscillation between successive left saddle-nodes \ref{fig:hybridconvectons}(e) and (g), the two central rolls first strengthen towards the right saddle-node \ref{fig:hybridconvectons}(f) before weakening, while a further pair of outer anticlockwise rolls nucleate and strengthen, thereby extending the convecton to four-rolls, as is typical for the growth mechanism along $L^-$ (cf. figure~\ref{fig:convecton_branch}(h)\textendash{}(j)).
The end rolls in the anticonvecton part of the multiconvecton undergo the breathing mechanism exhibited by anticonvectons along $\tilde{A}$ (cf. figure~\ref{fig:bdnsle5pr1n1_anticonvecton}(i)\textendash{}(k)), which results in the translation of the anticonvectons rolls away from and towards the end-walls at the right and left saddle-nodes, respectively.
Following $\tilde{L}^-$ between successive right saddle-nodes \ref{fig:hybridconvectons}(d) and (b) via the left saddle-node \ref{fig:hybridconvectons}(c), this breathing mechanism is accompanied by the anticonvecton part of the multconvecton oscillating between having a single roll at each end-wall at the right saddle-nodes and two rolls at each end-wall the left saddle-nodes, thereby replicating behavior seen along $\tilde{A}$ between saddle-nodes \ref{fig:bdnsle5pr1n1_anticonvecton}(h) and (i).

A similar analysis of multiconvectons $\tilde{L}^+$ can be carried out to reveal they are constituted of convectons $L^+$ and anticonvectons $\tilde{A}$ with similar growth mechanisms as for $\tilde{L}^-$.
Interpreting this branch poses more difficulty, however: domain size effects are quickly felt and the branch turns toward low energy right before saddle-node~(m). 
This is the reason why we did not provide any more figure panels for this side of the branch.

\section{Unbalanced systems with $N \neq -1$}\label{sec:Nneq1}

\subsection{Large-scale flow}\label{sec:lsf}

We now turn our attention to unbalanced regimes, i.e., $N \neq - 1$.
This does not affect the symmetry of the equations, which remain invariant with respect to $S_{\Delta}$.
Owing to the departure from $N= -1$, the basic state is no longer conductive but consists of a large-scale recirculating flow that develops from low Rayleigh numbers.
Basic properties of this large-scale flow can be ascertained by considering a parallel flow approximation similar to those applied in natural convection (e.g., \cite{batchelor1954heat}) or binary fluid convection in porous media (e.g., \cite{bahloul2003double}).
For this purpose, we consider an infinitely extended domain in the vertical direction and assume a purely conductive heat and solute transport between the side-walls, i.e., {\bf u} = {\bf 0} and ${T = C = x}$.
The buoyancy force in equation (\ref{eq:NS}) depends linearly on $T+NC=(1+N)x$.
It vanishes when $N=-1$ but exhibits linear horizontal dependence when ${N\neq -1}$.
As a result, when ${N > -1}$ (resp. ${N < -1}$), fluid at the right side-wall has an upward (resp. downward) buoyancy force, with the opposite at the left wall. 
This force imbalance increases when $N$ is driven away from $-1$.
The resulting horizontal buoyancy force gradients are a source of vorticity, which may be seen from the vorticity equation:
\begin{equation}
  \frac{1}{Pr}\left(\frac{\partial \omega}{\partial t} + \mathbf{u}\cdot\nabla \omega\right) = \nabla^2 \omega -Ra\left(\frac{\partial \Theta}{\partial x} + N \frac{\partial \Phi}{\partial x}\right) - Ra(1+N),
\end{equation}
where ${\omega =\hat{\mathbf{y}}\cdot\nabla \times\mathbf{u}}$ is the vorticity and ${\Theta = T-x}$ (resp. {${\Phi = C-x}$) is the deviation from the linear profile of temperature (resp. solutal concentration).
Assuming steady unidirectional flow in the vertical direction with ${\Theta = \Phi = 0}$, the vorticity equation reduces to
\begin{equation}
  \nabla^2 \omega = Ra(1+N)
\end{equation}
and the resulting vertical velocity profile is given by
\begin{equation}
  w(x,z) = -\frac{1}{24}Ra(1+N)\;x(2x-1)(2x+1), \label{eq:shear}
\end{equation}
resembling that found in natural convection \cite{batchelor1954heat}, except with a Rayleigh number that has been modified by the deviation of the buoyancy ratio away from ${N = -1}$.

\begin{figure}
  \centering
  \includegraphics[width=0.99\linewidth]{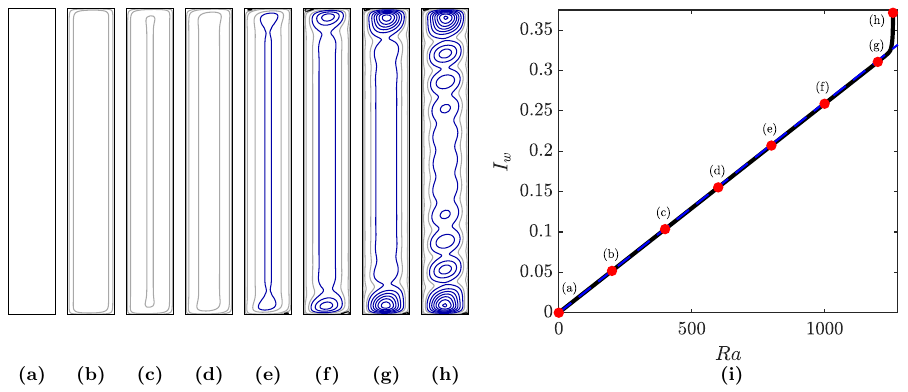}
  \caption[Development of the anticlockwise large-scale flow with $Ra$ when ${N = -0.9}$]{
    Development of the anticlockwise large-scale flow with $Ra$ until the first right saddle-node at $Ra \approx 1257$ when ${N = -0.9}$.
    (a)\textendash{}(h) Streamfunctions of the large-scale flow at the Rayleigh numbers marked in (i).
    Contours are shown using linear (blue) and logarithmic (black and grey) scales and take the values: $[-10^{-3}, -10^{-2}, -10^{-1}]$ (grey); $[10^{-3}, 10^{-2}, 10^{-1}]$ (black); $[-0.2,-0.4,-0.6,...]$ (blue).
    (i) Plot showing $I_w$ (\ref{eq:6_Iw}) for the large-scale flow at different Rayleigh numbers.
    The blue dashed line indicates the relationship ${I_w = \frac{1}{384}Ra(1+N)}$, obtained from equation (\ref{eq:shear}).}
  \label{fig:baseflowwmax}
\end{figure}

Figures~\ref{fig:baseflowwmax}(a)\textendash{}(h) show how this flow develops in the presence of end-walls in the vertical direction for ${N = -0.9}$.
We see that the strength of the flow increases with the Rayleigh number.
This is quantified in panel (i), which shows that the quantity
\begin{equation}
  I_w = \int_{0}^{0.5} w\left(x,z=\frac{L_z}{2}\right)\,dx, \label{eq:6_Iw} 
\end{equation} 
initially increases linearly with the Rayleigh number, in excellent agreement with the analytical prediction (\ref{eq:shear}), which implies ${I_w = \frac{1}{384}Ra(1+N)}$.

The calculation that led to expression (\ref{eq:shear}) is not valid near the horizontal end-walls of the domain due to the presence of no-slip boundary conditions.
Instead, we find that the vertical flow is forced sideways near the end-walls to generate a single domain-filling recirculating flow with anticlockwise circulation in thermally dominated flows (${N > -1}$) or clockwise circulation in solutally dominated flows (${N < -1}$).
Expression (\ref{eq:shear}) further breaks down towards larger Rayleigh numbers.
When ${N = -0.9}$, for example, we see that this breakdown first occurs via the formation of secondary rolls near the horizontal end-walls, as seen in figures~\ref{fig:baseflowwmax}(d)\textendash{}(g).
The origin of these rolls is doubly diffusive, which can be seen through the following argument in weakly thermally dominated flows.
As the fluid flows in an anticlockwise direction, hot and solute-rich fluid is drawn from the top right corner of the domain to the left.
This leftward moving flow is denser than the ambient fluid owing to the higher solute concentration and the fact that heat diffuses faster than solute.
As a result, it sinks again, near the left side-wall.
This fluid motion locally mixes the solute, which enhances the horizontal density gradient between side-walls.
This solute-rich fluid sinking along the left side-wall also generates negative vorticity, which, when sufficiently large, leads to an anticlockwise secondary roll at the top of the domain.
An analogous secondary roll forms in the bottom left corner of the domain owing to the fact that the large-scale flow preserves the $S_{\Delta}$ symmetry.
These rolls strengthen with increasing Rayleigh number and we see the formation of further secondary rolls near the first right saddle-node of the branch (figure~\ref{fig:bdnsle5pr1n1_anticonvecton}(h)), where their influence can be felt in the center of the domain, leading to the deviation of $I_w$ away from the infinite domain prediction when ${Ra > 1230}$.
We finally note that the formation of these secondary rolls, via doubly diffusive effects, differs from the initial formation of secondary rolls in natural convection \cite{vest1969stability}, which occurs via inertial effects.
We also found inertia-driven secondary rolls in this system; however, these were at considerably higher Rayleigh numbers, outside of the regime in which localized states are found.

\subsection{Small-amplitude unfolding}\label{sec:Nneq1_small}

In the balanced system we saw that the two branches of convectons, $L^+$ and $L^-$, bifurcate from a primary transcritical bifurcation of the conduction state at ${Ra \approx 1629.6}$ (figure~\ref{fig:pribifN1}).
This does not occur in the unbalanced systems as the primary transcritical bifurcation unfolds when ${N \neq -1}$ due to the presence of the large-scale flow.
To understand what happens to convectons at large amplitude, we therefore start by exploring this unfolding at small amplitude and extend earlier results by Bardan et al. \cite{bardan2000nonlinear}, who considered two values of $N$ in small domains.

\begin{figure}
  \centering
  \includegraphics[width=0.99\linewidth]{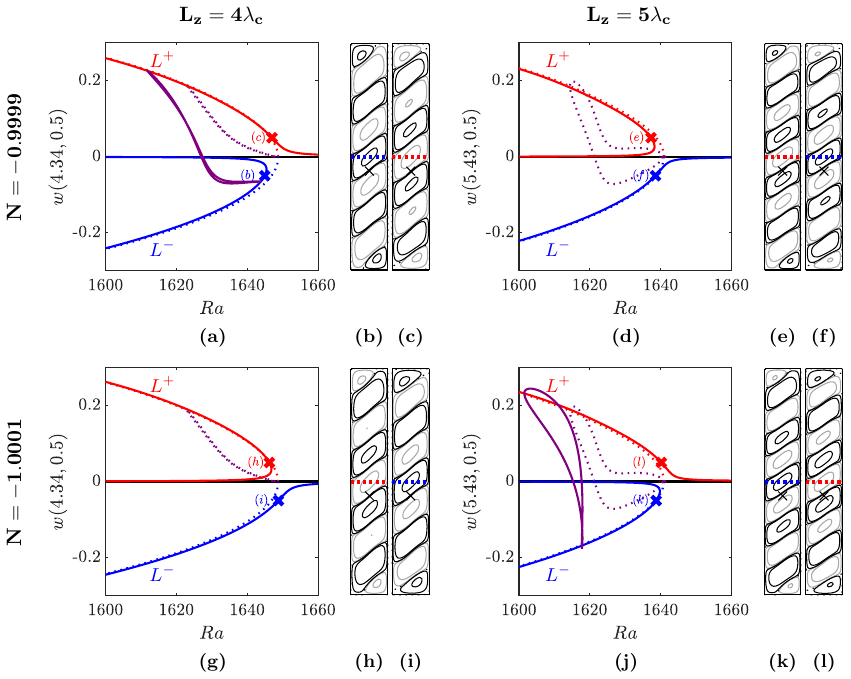}
  \caption[Unfolding of the primary transcritical bifurcation in small domains of length ${L_z = 4\lambda_c}$ and ${L_z = 5\lambda_c}$ for ${N = -0.9999}$ and ${N = -1.0001}$]{Unfolding of the primary transcritical bifurcation in small domains of length ${L_z = 4\lambda_c}$ (left set of panels) and ${L_z = 5\lambda_c}$ (right set of panels) for ${N = -0.9999}$ (top row) and ${N = -1.0001}$ (bottom row).
    (a), (d), (g), (j) Bifurcation diagrams showing the vertical velocity of states at the point marked by black crosses in the streamfunction plots against Rayleigh number for branches with ${N=-1}$ (dotted lines) and either (a), (d) ${N=-0.9999}$ or (g), (j) ${N = -1.0001}$ (solid lines).
    The colors red, blue and purple indicate $L^+$, $L^-$ and branches of asymmetric states, respectively.
    The streamfunction plots are shown at the marked points in the bifurcation diagrams. 
    The horizontal midline is shown using a red or blue dotted line to indicate whether the state lies on $L^-$ (blue) or $L^+$ (red).
    The contour values used were: $[-10^{-3}, -10^{-2}, -10^{-1}]$ (grey); and $[10^{-3}, 10^{-2}, 10^{-1}]$ (black).
  }
  \label{fig:unfolding_explanation}
\end{figure}
\subsubsection{Small domains}
As in the balanced case with ${N = -1}$, we first consider smaller domains whose vertical extent is either ${L_z = 4\lambda_c}$ (left panels in figure~\ref{fig:unfolding_explanation}) or ${L_z = 5\lambda_c}$ (right set of columns in figure~\ref{fig:unfolding_explanation}) where the unfolding of primary bifurcations is clearer.
Regardless of the buoyancy ratio and domain size, the branches containing states with a clockwise central roll (black in the streamfunction plots in figure~\ref{fig:unfolding_explanation}) extend to large amplitude as the $L^-$ branch (blue), while those with an anticlockwise central roll extend to larger amplitudes as $L^+$ (red).

Figure~\ref{fig:unfolding_explanation} indicates how the unfolding of the primary transcritical bifurcation varies with both domain size and buoyancy ratio.
In particular, when ${N = -0.9999}$, we find that the branch of large-scale flow developing from ${Ra = 0}$ connects to the branch initially associated with the states containing weak clockwise rolls in the top left and bottom right corners of the domain, i.e., $L^-$ when ${L_z = 4\lambda_c}$ (figure~\ref{fig:unfolding_explanation}(b)) and $L^+$ when ${L_z = 5\lambda_c}$ (figure~\ref{fig:unfolding_explanation}(e)).
Such behavior likely arises because the anticlockwise roll inclined downwards from the top right corner of the domain (or the symmetry-related roll in the bottom left corner) reinforces the large-scale flow with the same sense of rotation, while the weaker clockwise corner rolls form via viscous effects.
Since the number of pairs of rolls in the eigenmode associated with the transcritical bifurcation when ${N=-1}$ varies with the domain size, the branch containing states with clockwise corner rolls also changes and, consequently, so does the unfolding of the transcritical bifurcation.
The results when ${N = -1.0001}$ are analogous except that we find that the branch of large-scale flow developing from ${Ra = 0}$ now connects to the branch initially associated with states containing weak anticlockwise rolls in the top left and bottom right corners of the domain, i.e., $L^+$ when ${L_z = 4\lambda_c}$ (figure~\ref{fig:unfolding_explanation}(h)) and $L^-$ when ${L_z = 5\lambda_c}$ (figure~\ref{fig:unfolding_explanation}(k)).

In figure~\ref{fig:N1_Lz4Lz5}, we saw that when ${N = -1}$ the conduction state undergoes a pitchfork bifurcation prior to the transcritical bifurcation and the resulting branches (purple dotted lines in figure~\ref{fig:unfolding_explanation}) terminate at a pitchfork bifurcation on $L^+$ (red) for both ${L_z \approx 4\lambda_c}$ and ${L_z \approx 5\lambda_c}$.
Unlike the transcritical bifurcation considered above, this pitchfork bifurcation does not unfold as the buoyancy ratio varies away from ${N=-1}$ because the large-scale flow preserves the $S_{\Delta}$ symmetry of the system \cite{bardan2000nonlinear}. 
Nevertheless, this bifurcation and the bifurcating branches are affected both by the buoyancy ratio and the domain size, as illustrated by the presence or absence of purple lines in figure~\ref{fig:unfolding_explanation}. 
When $N$ departs from $-1$, the primary pitchfork bifurcation is transferred to the branch that originates from ${Ra =0}$ ($L^-$ (blue) when ${L_z = 4\lambda_c}$, ${N>-1}$ or ${L_z = 5\lambda_c}$, ${N <-1}$; $L^+$ (red) when ${L_z = 4\lambda_c}$, ${N <-1}$ or ${L_z = 5\lambda_c}$, ${N >-1}$), but the termination point of the resulting branches remains on $L^+$.
This leads to short branch segments that either connect $L^-$ to $L^+$ or connect $L^+$ to itself.
The former connections persist as $N$ is varied away from the balanced case, as evidenced in figures~\ref{fig:unfolding_explanation}(a) and (j).
In the latter case, however, the two pitchfork bifurcations approach each other on $L^+$ and ultimately collide (at ${N\approx -0.999902}$ when ${L_z = 5\lambda_c}$ and ${N\approx -1.00009}$ when ${L_z = 4\lambda_c}$), so that the aforementioned pitchfork bifurcations are absent from the bifurcation diagrams presented in figures~\ref{fig:unfolding_explanation}(d) and (g).

\subsubsection{Large domains}

\begin{figure}
  \centering
  \includegraphics[width=0.99\linewidth]{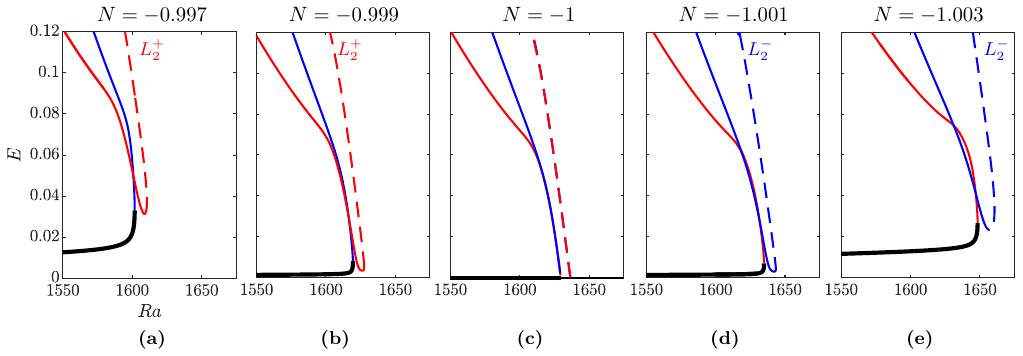}
  \caption[Unfolding of the primary bifurcation in a domain with ${L_z =12\lambda_c}$ for $N$ between ${N = -0.997}$ and ${N=-1.003}$]{Unfolding of the primary bifurcation in a domain with ${L_z =12 \lambda_c}$ for $N$ between (a) ${N = -0.997}$ and (e) ${N=-1.003}$.
    The branch segments shown include the conduction state or large-scale flow (black), $L^-$ (blue), $L^+$ (red) and the branch of two-pulse states, $L^+_2$ (red dashed) or $L^-_2$ (blue dashed), that connects to one of the convecton branches.}
  \label{fig:N1003_N0997}
\end{figure}
Since we are primarily interested in how convectons are affected by variations in the buoyancy ratio, we consider the larger domain: ${L_z=12\lambda_c}$.
The unfolding of the primary transcritical bifurcation is illustrated in figure~\ref{fig:N1003_N0997}, where we see similarities to the unfolding in the smaller domain with ${L_z=4\lambda_c}$ (figures~\ref{fig:unfolding_explanation}(a) and (g)).
In particular, we find that the branch of large-scale flow originating from low Rayleigh numbers connects to $L^-$ when ${N > -1}$ and to $L^+$ when ${N < -1}$.
As with the smaller domains, the second convecton branch connects at small amplitude to the branch of large-scale flow states following the primary transcritical bifurcation.
However, this behavior is only observable over a small range of Rayleigh numbers in figure~\ref{fig:N1003_N0997} since this branch segment proceeds to connect to a branch of two-pulse convectons originating from the unfolding of the second transcritical bifurcation.
Here, we find that $L^+$ connects to $L^+_2$ when ${N > -1}$, while $L^-$ connects to $L_2^-$ when ${N < -1}$.

Details of the unfolding are further seen in figure~\ref{fig:N0999_small}, which shows an enlargement of the bifurcation diagram for ${N = -0.999}$ (figure~\ref{fig:N1003_N0997}(b)), together with selected states along each branch segment.
As the branch of weak anticlockwise base flow states (e.g., figure~\ref{fig:N0999_small}(b)) is followed towards larger Rayleigh numbers, the single large roll breaks up into a series of weak counterrotating rolls with a clockwise roll in the center of the domain (figure~\ref{fig:N0999_small}(c)).
The central rolls proceed to strengthen along $L^-$, while the outer rolls weaken and are replaced by a pair of weak anticlockwise flows that extend vertically from the central rolls to either vertical end of the domain (figure~\ref{fig:N0999_small}(d)).

Convectons on $L^+$ at small amplitude assume a similar form with a central anticlockwise roll being immediately surrounded by weak counterrotating rolls and then by a pair of large anticlockwise rolls that fill the remainder of the domain (figure~\ref{fig:N0999_small}(e)).
The large outer rolls break up as the branch is followed towards states with lower kinetic energy that originated from the conduction state (figure~\ref{fig:N0999_small}(f)).
The central rolls weaken during this period and continue to do so while other interior rolls strengthen to give multi-pulse states on $L^+_2$ as this branch turns back towards lower Rayleigh numbers (figure~\ref{fig:N0999_small}(g)).
The subsequent behavior of this two-pulse convecton branch appears to be sensitive to the domain size considered here since we find that the branch turns back on itself and re-enters the parameter regime depicted in figure~\ref{fig:N0999_small}(a), before leaving by following the branch segment associated with the second two-pulse convecton branch, $L_2^-$, (figure~\ref{fig:N0999_small}(h)). 
This process is associated with repositioning of rolls within the states and occurs several times, as can be seen from the three saddle-nodes within the range ${1590<Ra<1600}$ in figure~\ref{fig:N0999_small}.
The branch finally exits this small-amplitude region and snakes towards larger amplitude states at lower Rayleigh numbers following $L_2^-$.

\begin{figure}
  \centering
  \includegraphics[]{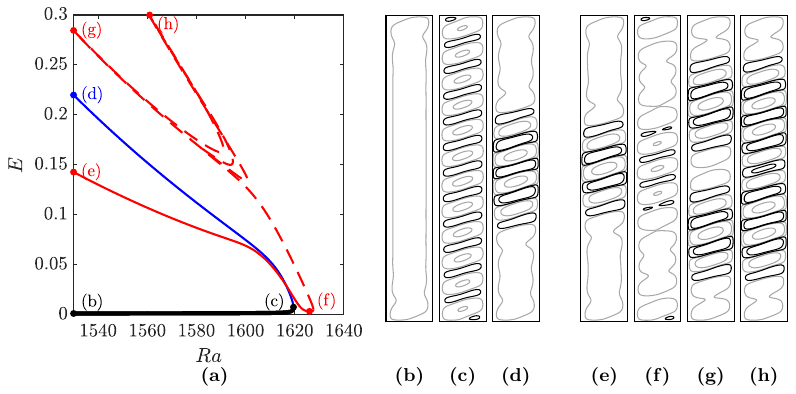}
  \caption[Bifurcation diagram and streamfunctions for the small-amplitude behavior when ${N = -0.999}$]{(a) Extension of the bifurcation diagram for ${N = -0.999}$ in a domain with $L_z =12\lambda_c$ shown in figure~\ref{fig:N1003_N0997}(b).
    (b)\textendash{}(h) Streamfunctions of the small-amplitude profiles along the different branch segments.
    The contour values used were: $[-10^{-3}, -10^{-2}, -10^{-1}]$ (grey); and $[10^{-3}, 10^{-2}, 10^{-1}]$ (black).
  }
  \label{fig:N0999_small}
\end{figure}


\subsection{Convectons with varying buoyancy ratio in the thermally dominated regime}\label{sec:conv_Nneq1}

We restrict our attention to convectons in thermally dominated regimes ($N > -1$).
As the large-scale flow and the convecton rolls have the same sense of rotation, we expect the results to be simpler than the solutally dominated regimes ($N < -1$).

\begin{figure}
  \centering
  \includegraphics[width=0.99\linewidth]{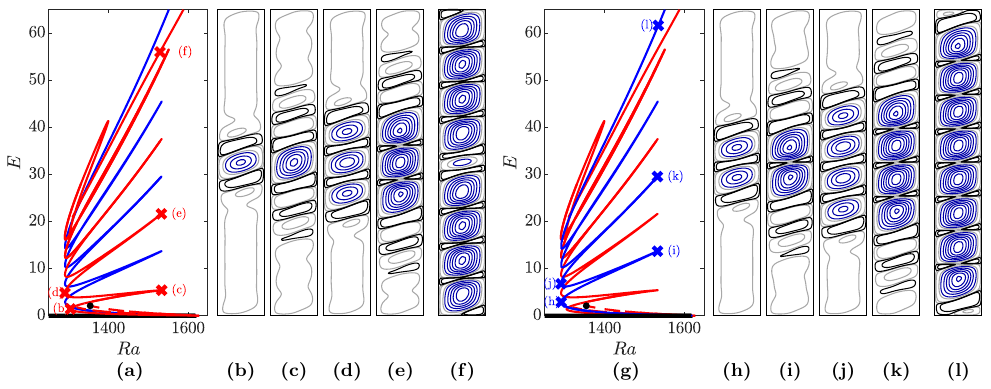}
  \caption[Convecton branches and streamfunction profiles for ${N = -0.999}$, ${Pr = 1}$ and ${Le = 5}$ ]{Convecton branches and streamfunction profiles for ${N = -0.999}$, ${Pr = 1}$ and ${Le = 5}$. 
    (a) and (g) Bifurcation diagram showing the convecton branches $L^+$ (red) and $L^-$ (blue) and a branch of two-pulse states (red dashed) that is terminated at small-amplitude at the black dot.
    (b)\textendash{}(f), (h)\textendash{}(l) Streamfunctions of: (b)\textendash{}(e) convectons at four saddle-nodes of $L^+$ that are marked in (a);
    (f) the domain-filling state on $L^+$ marked in (a);
    (h)\textendash{}(k) convectons at four saddle-nodes of $L^-$ marked in (g);
    (l) the domain-filling state on $L^-$ marked in (g).
    Contours are shown using linear (blue) and logarithmic (black and grey) scales and take the values: $[-10^{-3}, -10^{-2}, -10^{-1}]$ (grey); $[10^{-3}, 10^{-2}, 10^{-1}]$ (black); $[-0.2,-0.4,-0.6,...]$ (blue).
  }	\label{fig:N0999}
\end{figure}
In weakly unbalanced systems, the branches involved in the above unfolding (figure~\ref{fig:N1003_N0997}) proceed to undergo homoclinic snaking as they extend towards larger amplitudes.
This behavior is illustrated in figure~\ref{fig:N0999} for ${N = -0.999}$, where we see the close resemblance of the snaking branches to those in the balanced system for ${N = -1}$ (figure~\ref{fig:convecton_branch}).
The strong central anticlockwise rolls in the convectons (figures~\ref{fig:N0999}(b)\textendash{}(e), \ref{fig:N0999}(h)\textendash{}(k)) and the domain-filling states (figures~\ref{fig:N0999}(f), (l)) that lie on these branches are also minimally affected by the small change to the buoyancy ratio.

The imbalance in the system can, however, be seen in the large, weak rolls that fill the domain outside of the central rolls and whose origin can be traced back to the large-scale flow considered in section~\ref{sec:lsf}.
These rolls are sensitive to increases in the Rayleigh number, with each of them breaking up into smaller counterrotating rolls between successive left and right saddle-nodes (e.g., compare figures~\ref{fig:N0999}(d) and (e)) as they cannot overcome the strengthening of the rolls that contribute to the front of the convecton with increasing Rayleigh number.
This breakup allows the pair of weak anticlockwise rolls closest to the central rolls to strengthen between right and left saddle-nodes (e.g., figures~\ref{fig:N0999}(c)\textendash{}(d) and (i)\textendash{}(j)), which increases the number of strong anticlockwise rolls by two over each snaking oscillation.
This behavior appears similar to that observed for $N=-1$ (figure~\ref{fig:convecton_branch}).


\subsection{Effect of anticonvecton branch on convectons}\label{sec:anti_Nneq1}

\begin{figure}
  \centering
  \includegraphics[width=0.99\linewidth]{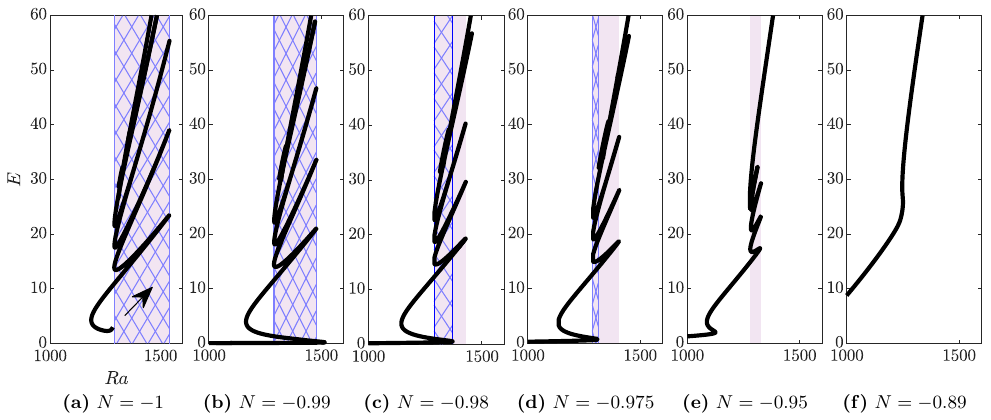}
  \caption[The anticonvecton branch for ${N = -1}$, ${N -0.99}$, ${N =-0.98}$, ${N =-0.975}$, ${N =-0.95}$ and ${N=-0.89}$]{The anticonvecton branch for: (a) ${N = -1}$, (b) ${N -0.99}$, (c) ${N =-0.98}$, (d) ${N =-0.975}$, (e) ${N =-0.95}$ and (f) ${N=-0.89}$.
    The anticonvecton pinning region is shaded in light purple, while the convecton pinning region is hatched in blue when it exists.
    The black arrow in (a) indicates the direction in which the anticonvecton branch continues when ${N =-1}$ (cf. figure~\ref{fig:bdnsle5pr1n1_anticonvecton}).
  }
  \label{fig:anticonvectonN}
\end{figure}
The bifurcation diagram for $N=-0.9$ (Figure \ref{fig:baseflowwmax}) hints at the fact that convectons cease to exist in the thermally dominated regime of doubly diffusive convection.
To understand this further, we first notice that the convecton pinning region depends upon the structure of the anticonvecton branch.
This dependence is illustrated in figure~\ref{fig:anticonvectonN}, where the bifurcation diagrams for selected $N$ show the anticonvecton branch (black) and the pinning region associated with its snaking (light purple), together with the convecton pinning region (blue hatched).
This figure shows that the convecton pinning region, when it exists, is bounded on the left by the lower edge of the anticonvecton pinning region.
In contrast, the right edge of the convecton pinning region changes with increasing $N$, owing to the formation of a small-amplitude right saddle-node at ${N \approx -0.9904}$, when the anticonvecton branch connects to the branch of large-scale flow originating from ${Ra = 0}$, and its subsequent motion towards lower Rayleigh numbers. 
Initially, the right edge of the convecton pinning region coincides with the upper edge of the anticonvecton pinning region (figures~\ref{fig:anticonvectonN}(a) and (b)), but changes to being bounded by the first right saddle-node of the anticonvecton branch when this saddle-node is located within the anticonvecton pinning region (figure~\ref{fig:anticonvectonN}(c)).
Consequently, the convecton pinning region disappears (e.g., figures~\ref{fig:anticonvectonN}(d)\textendash{}(f)) when this right saddle-node passes through the left edge of the anticonvecton pinning region.

\begin{figure}
  \centering
  \includegraphics[]{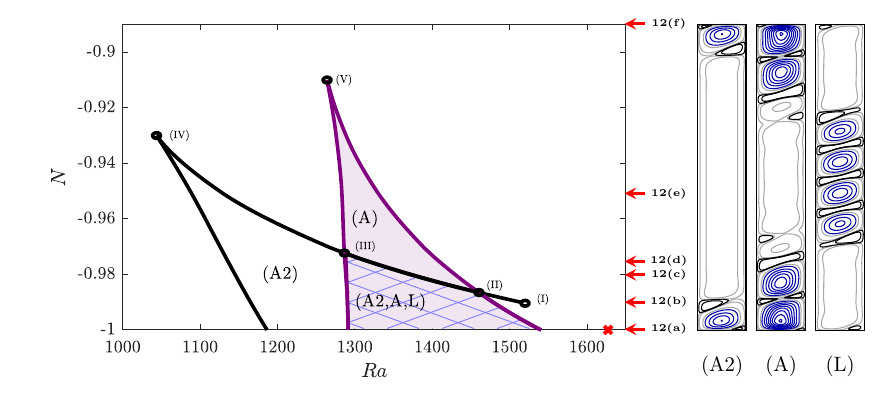}
  \caption[Regions of $(Ra,N)$ parameter space in which two-roll anticonvectons (A2), snaking anticonvectons (A) and convectons (L) can be found]{Regions of $(Ra,N)$ parameter space in which two-roll anticonvectons (A2), snaking anticonvectons (A), shaded in light purple, and convectons (L), hatched in blue, can be found.
    The boundaries between regions are given by: the first left and right saddle-nodes of the anticonvecton branch (black) and the pinning region of the snaking anticonvectons (purple).
    Five codimension-two points where the boundaries intersect are labelled and marked by the black circles.
    The red cross marks the primary bifurcation of the conduction state when ${N = -1}$.
    The three panels on the right give representative examples of the three types of states.	
    Arrows indicate the values of $N$ used to produce the bifurcation diagrams in figure~\ref{fig:anticonvectonN}.}
  \label{fig:pinning_summary}
\end{figure}

Figure~\ref{fig:pinning_summary} summarises the above changes in $(Ra,N)$ parameter space by tracking the following over a range of $N$: the convecton pinning region (blue hatched), in which convectons (L) exist; the anticonvecton pinning region (light purple), in which anticonvectons (A) exist; and the first left saddle-node of the anticonvecton branch and, when it exists, the first right saddle-node of the anticonvecton branch.
We call the lowest part of the anticonvecton branch, to the left of the anticonvecton pinning region and between the first left and right saddle-nodes, the two-roll anticonvecton (A2) branch, with one roll at each end-wall.
From this figure, we identify five codimension-two points where saddle-nodes corresponding to the boundaries either originate (I), cross through each other (II) and (III) or collide and disappear (IV) and (V).
The qualitative structure of the bifurcation diagrams changes near these points, as seen in figure~\ref{fig:anticonvectonN}.
We discuss these changes in the remainder of this section.

\subsubsection{(I): Changes at small-amplitude}

The small-amplitude structure of the bifurcation diagram is the first to change as the buoyancy ratio is increased into the thermally dominated regime.
We find that the behavior changes from the unfolding discussed in figure~\ref{fig:N1003_N0997}, where the branch of large-scale flow originating from the origin connects to $L^-$ before extending towards larger amplitudes, to involve branches of states with weak anticlockwise rolls at both ends of the domain.
This occurs in several stages and involves intermediate branches that we will not consider in detail.
For simplicity, we focus on two aspects of the small-amplitude reconnections around $N\approx -0.9904$ (point (I) in figure~\ref{fig:pinning_summary}): firstly, how the anticonvecton branch connects to the branch of large-scale flow (figure~\ref{fig:trackanticonvectons}) and, secondly, which branch $L^-$ connects to instead (figure~\ref{fig:trackconvectons}).

\begin{figure}
  \centering
  \includegraphics[width=0.99\linewidth]{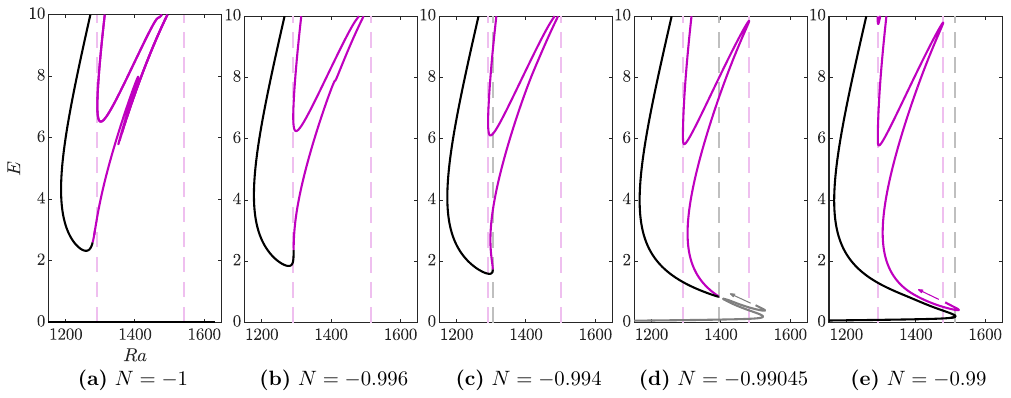}
  \caption[Bifurcation diagrams showing how the anticonvecton branch $A$ disconnects from $\tilde{A}$ and instead connects to the branch of base flow between ${N = -1}$ and ${N = -0.99}$]{Bifurcation diagrams showing how the anticonvecton branch $A$ (black) disconnects from $\tilde{A}$ (purple) and instead connects to the branch of base flow between ${N = -1}$ and ${N = -0.99}$. 
    Arrows indicate the direction in which the branches continue.
    The purple vertical dashed lines indicate the limits of the anticonvecton pinning region. 
    The grey vertical dashed lines in (c)\textendash{}(e) indicate the location of the lower right saddle-node of the anticonvecton branch.}
  \label{fig:trackanticonvectons}
\end{figure}

Figure~\ref{fig:trackanticonvectons} depicts how the lower sections of the anticonvecton branches change when $N$ is increased from ${N=-1}$ (figure~\ref{fig:trackanticonvectons}(a)) to ${N=-0.99}$ (figure~\ref{fig:trackanticonvectons}(e)).
This branch segment undergoes a cusp bifurcation when ${N\approx-0.996}$, which leads to the two new saddle-nodes seen in figure~\ref{fig:trackanticonvectons}(c) when ${N=-0.994}$. 
The right saddle-node of this pair extends towards lower energies and larger Rayleigh numbers as $N$ increases and approaches the left saddle-node that an intermediate branch (grey in figure~\ref{fig:trackanticonvectons}(d)) undergoes before snaking towards larger amplitudes, which is indicated by the upward-pointing grey arrow.
The two saddle-nodes connect in a transcritical bifurcation around ${N\approx -0.9904}$ and the two branches of anticonvectons separate as this bifurcation unfolds.
This results in the branch of large-scale flow connecting to $A$ when ${N=-0.99}$, while $\tilde{A}$ connects with the intermediate branch and continues towards larger amplitudes (figure~\ref{fig:trackanticonvectons}(e)).

\begin{figure}
  \centering
  \includegraphics[width=0.99\linewidth]{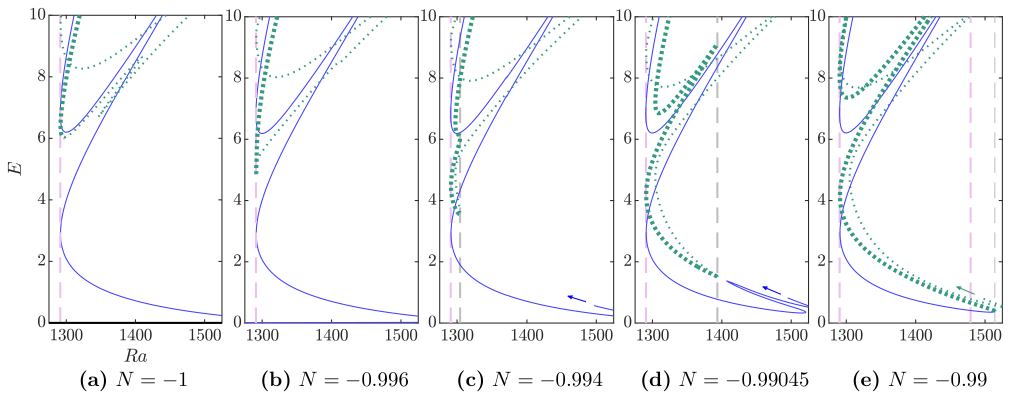}
  \caption[Bifurcation diagrams showing how the convecton branch $L^-$ connects to the branch of multiconvectons $\tilde{L}^-$ between ${N = -1}$ and ${N = -0.99}$]{Bifurcation diagrams showing how the convecton branch $L^-$ (blue) connects to the branch of multiconvectons $\tilde{L}^-$ (green dotted) between ${N = -1}$ and ${N = -0.99}$. 
    Arrows indicate the direction in which the branches continue.
    As in figure~\ref{fig:trackanticonvectons}, the purple vertical dashed lines indicate the limits of the anticonvecton pinning region, while the grey vertical dashed lines in (c)\textendash{}(e) indicate the location of the lower right saddle-node of the anticonvecton branch. 
  }
  \label{fig:trackconvectons}
\end{figure}

Other branches also extend towards smaller amplitudes as the buoyancy ratio increases, including branches of multiconvectons that consist of states containing central rolls (like convectons) and an anticlockwise roll at either end of the domain (see figures~\ref{fig:N099}(b.ii)\textendash{}(b.v) and (c.ii)\textendash{}(c.v) for examples).
Figure~\ref{fig:trackconvectons} illustrates this behavior for the branch of even multiconvectons $\tilde{L}^-$ (green dotted) and how this leads to the branch connecting with $L^-$ (blue) by ${N = -0.99}$ (figure~\ref{fig:trackconvectons}(e)).
In the balanced system (figure~\ref{fig:trackconvectons}(a), but also figure \ref{fig:hybridconvectons}(a)), this multiconvecton branch snakes upwards in both directions from the left saddle-node at ${E\approx 6}$, as indicated by the thick and thin green dotted lines.
As $N$ increases, this lower left saddle-node moves towards lower energies and the adjacent branch segments undergo a cusp bifurcation around ${N\approx -0.996}$, which leads to the additional saddle-nodes seen on the lower part of $\tilde{L}^-$ by ${N= -0.994}$ (figure~\ref{fig:trackconvectons}(c)).
The newly formed right saddle-nodes are located at similar values of the Rayleigh number as the right saddle-node that forms on the anticonvecton branch in figure~\ref{fig:trackanticonvectons}, as evidenced by the dashed vertical grey lines in both figures.
This continues to be the case as $N$ increases to ${N=-0.99045}$ (figure~\ref{fig:trackconvectons}(d)).
Meanwhile, we find that $L^-$ disconnects from the branch of large-scale flow between ${N=-0.996}$ and ${N=-0.994}$ and, instead, continues by snaking upwards (see blue arrows in figure~\ref{fig:trackconvectons}).
Similarly to the intermediate branch previously discussed, $L^-$ exhibits a left saddle-node that approaches the right saddle-node of $\tilde{L}^-$ when ${N=-0.99045}$ (not shown) before connection at around ${N\approx -0.9904}$.
The branch of odd multiconvectons $\tilde{L}^+$ exhibits analogous changes and connects to $L^+$ by ${N=-0.99}$.

\begin{figure}
  \flushleft
  \includegraphics[width=0.99\linewidth]{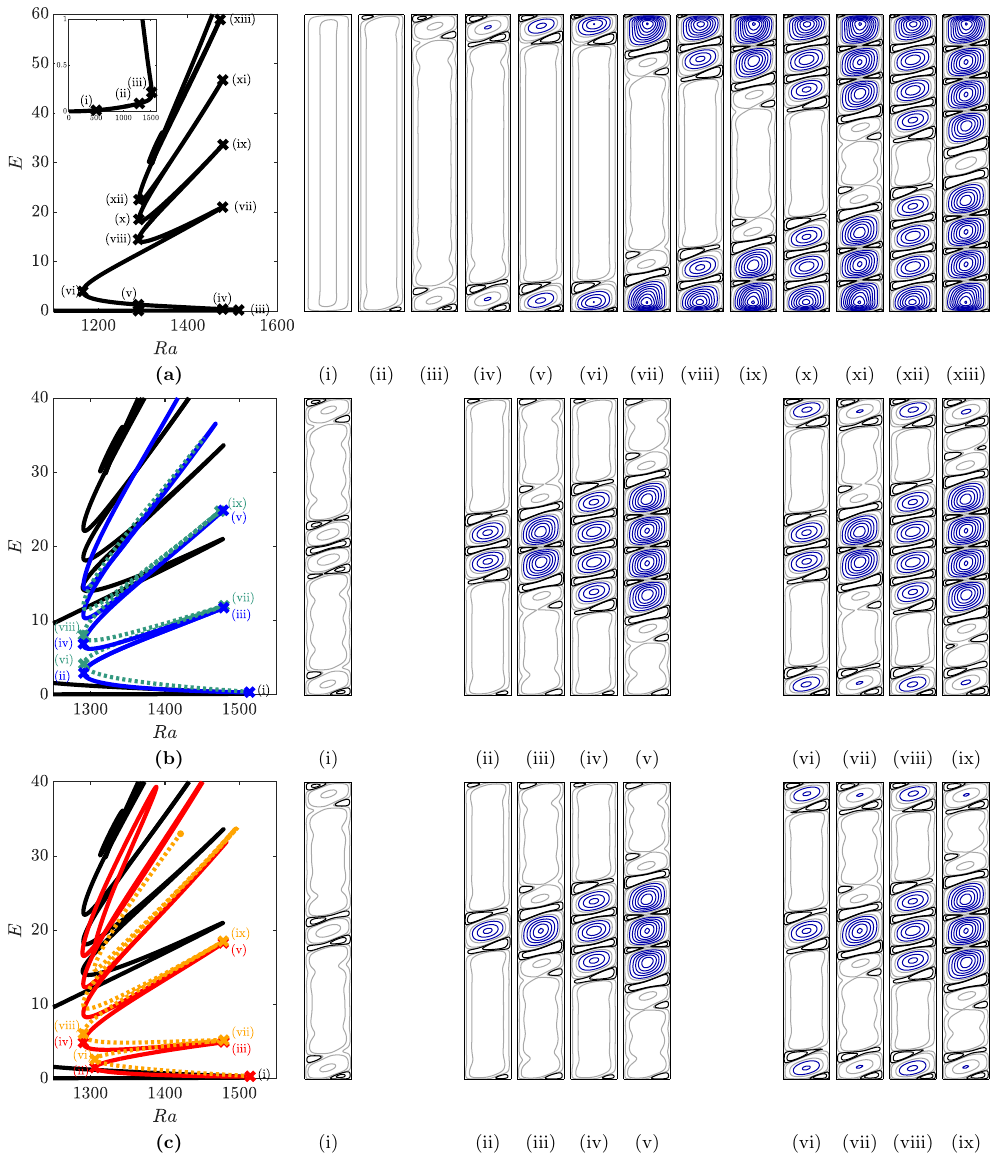}
  \caption[Bifurcation diagrams and streamfunctions for steady states when ${N = -0.99}$]{\scriptsize Bifurcation diagrams and streamfunctions for steady states when ${N = -0.99}$.
    Top row: (a) Branch of anticonvectons $A$ (black) and (i)\textendash{}(xiii) selected profiles of anticonvectons.
    Middle row: (b) Branch of even convectons $L^-$ (blue) and even multiconvectons $\tilde{L}^-$ (green); (i) State at the saddle-node where $L^-$ and $\tilde{L}^-$ join; States at the first four saddle-nodes of (ii)\textendash{}(v) $L^-$ and (vi)\textendash(ix) $\tilde{L}^-$
    Bottom row: (b) Branch of odd convectons $L^+$ (red) and odd multiconvectons $\tilde{L}^+$ (orange); (i) State at the saddle-node where $L^+$ and $\tilde{L}^+$ join; States at the first four saddle-nodes of (ii)\textendash{}(v) $L^+$ and (vi)\textendash(ix) $\tilde{L}^+$.
    Contours are shown using linear (blue) and logarithmic (black and grey) scales and take the values: $[-10^{-3}, -10^{-2}, -10^{-1}]$ (grey); $[10^{-3}, 10^{-2}, 10^{-1}]$ (black); $[-0.2,-0.4,-0.6,...]$ (blue).}
  \label{fig:N099}
\end{figure}

Figure~\ref{fig:N099} depicts sections of the bifurcation diagrams and streamfunctions of states along the branches for ${N=-0.99}$ after these reconnections have occurred and the branch of anticonvectons connects to the branch of large-scale flow (top row), while each branch of convectons connects to the corresponding branch of multiconvectons (bottom two rows). 
The initial behavior of the anticonvecton branch is similar to that previously seen when ${N=-0.9}$ (figure~\ref{fig:baseflowwmax}), as a domain-filling anticlockwise flow develops across the domain and strengthens with increasing Rayleigh number (figures~\ref{fig:N099}(a.i) and (a.ii)).
This branch turns around at the right saddle-node marked (iii), where the state has one secondary anticlockwise roll at either end of the domain. 
These rolls strengthen as the branch is followed towards the following left saddle-node marked (vi) and separate from the weak domain-filling flow as the small clockwise rolls, first seen in (a.iii), strengthen and extend across the full horizontal extent of the domain (figures~\ref{fig:N099}(a.iii)\textendash{}(a.vi)).
The subsequent behavior of this anticonvecton branch closely resembles that when ${N=-1}$ (figure~\ref{fig:bdnsle5pr1n1_anticonvecton}), where the anticonvectons extend by a pair of strong anticlockwise rolls across each oscillation as the branch snakes upwards until the domain has nearly filled (a.xiii).
The main difference, however, is the presence of the weak anticlockwise flow that fills the region between the two sets of stronger rolls. 

As a result of the reconnection process illustrated in figure~\ref{fig:trackconvectons} and the corresponding one for $L^+$ and $\tilde{L}^+$, we find that each branch of convectons (blue and red solid lines) connects with the corresponding branch of multiconvectons (green and orange dotted lines) at a small-amplitude right saddle-node, as shown in figures~\ref{fig:N099}(b) and (c).
Figures~\ref{fig:N099}(b.i) and (c.i) depict the associated state, which we see contain either one (c.i) or two (b.i) weak central anticlockwise roll(s) separated by a pair of weak clockwise rolls from the larger anticlockwise rolls, which each contain a secondary anticlockwise end roll with amplitude comparable to the central roll(s).
After these right saddle-nodes, each of the convecton and multiconvecton branches proceed to snake upwards within the anticonvecton pinning region before finite domain effects become important at large amplitudes. 
During this upwards snaking process, the central rolls of both types of states develop in similar ways (compare figures~\ref{fig:N099}(b.ii)\textendash{}(b.v) to (b.vi)\textendash{}(b.ix) and (c.ii)\textendash{}(c.v) to (c.vi)\textendash{}(c.ix)).
As the branches are followed between successive left and right saddle-nodes, the central anticlockwise rolls strengthen and a secondary roll develops on the interior side of each of the large, weak anticlockwise rolls. 
The central rolls proceed to weaken while the secondary rolls continue to strengthen and disconnect from the background flow as each branch is followed towards the following left saddle-node, resulting in the increase of the number of rolls of the central structure by two over a single snaking oscillation.

Differences between convectons and multiconvectons, however, are observed by the presence or absence of strong anticlockwise rolls located at either end of the domain.
Starting from the right saddle-nodes where the corresponding branches meet ((i) in figures~\ref{fig:N099}(b) and (c)) and heading towards lower Rayleigh numbers along either convecton branch, the pair of anticlockwise end rolls seen in figures~\ref{fig:N099}(b.i) and (c.i) weaken.
This leads to each of the large background rolls extending from the central rolls to one end of the domain and the absence of stronger anticlockwise end rolls in convectons.
In contrast, as each multiconvecton branch is followed from its small-amplitude right saddle-node, marked (i) in figures~\ref{fig:N099}(b) and (c), towards lower Rayleigh numbers, the secondary anticlockwise end rolls strengthen and separate from the large background rolls, as seen in figures~\ref{fig:N099}(b.vi) and (c.vi).
The strength of these end rolls is governed by the segment (iv)\textendash{}(v) on the anticonvecton branch, as can be seen by comparing these end rolls at the left saddle-nodes ((b.vi), (b.viii), (c.vi) and (c.viii)) to (a.v) and those at the right saddle-nodes ((b.vii), (b.ix), (c.vii) and (c.ix)) to (a.iv). 
Thus, in contrast to the central rolls of both convectons and multiconvectons that strengthen with increasing Rayleigh number, the end rolls in the multiconvectons weaken with increasing Rayleigh number.

\subsubsection{(II): From snaking to isolas}

The snaking of the connected branches of convectons and multiconvectons seen in figure~\ref{fig:N099} persists until ${N\approx -0.9866}$ (point (II) in figure~\ref{fig:pinning_summary}), when the first right saddle-node of the anticonvecton branch (figure~\ref{fig:N099}(a.iii)) coincides with the right edge of the convecton and anticonvecton pinning regions.
As this right saddle-node passes into the anticonvecton pinning region for larger values of the buoyancy ratio, the snaking branches break up into a set of vertically stacked isolas that are bounded between this saddle-node and the left edge of the anticonvecton pinning region, as exemplified in figures~\ref{fig:bdn0975}(a) and (b).
This is the same process that was termed `Snake death II: Isola formation' by Yulin and Champneys \cite{yulin2010discrete}, who first observed it in a model of a discrete optical cavity with detuning and a linear loss term.
Champneys et al. \cite{champneys2012homoclinic} later presented a theoretical argument explaining how such behavior can arise from the unfolding of a saddle-center bifurcation when a fold in a branch of homogeneous states coincides with an edge of the pinning region.
This appears to also be the case here in the doubly diffusive system, where the anticonvecton branch at small amplitude acts as the branch of homogeneous states.

The manner in which the snaking branches break up into a set of vertically stacked isolas can be understood by considering the unfolding of transcritical bifurcations that occur between the convecton and multiconvecton branches.
Figure~\ref{fig:snakestoisola} depicts one such example near the right saddle-nodes associated with convectons and multiconvectons with four central rolls. 
The right saddle-nodes of $L^-$ and $\tilde{L}^-$ approach as the buoyancy ratio increases between ${N=-0.98664}$ (figure~\ref{fig:snakestoisola}(a)) and ${N=-0.98663}$ (figure~\ref{fig:snakestoisola}(b)), before connecting in a transcritical bifurcation between ${N=-0.98663}$ and ${N=-0.98662}$ (figure~\ref{fig:snakestoisola}(c)).
As this bifurcation unfolds upon further increase of the buoyancy ratio, the upper and lower branch segments of $L^-$ separate and respectively connect to the upper and lower branch segments of $\tilde{L}^-$, as seen in figures~\ref{fig:snakestoisola}(c) and (d).
This leads to a pair of branches connecting convectons to multiconvectons.
While it has not been confirmed here, it is likely that right saddle-nodes associated with convectons with different number of central rolls reconnect over a small range of buoyancy ratios, which would result in an intermediate combination of isolas and snaking, similar to behavior that Yulin and Champneys \cite{yulin2010discrete} found.

\begin{figure}
  \centering
  \includegraphics[width=0.99\linewidth]{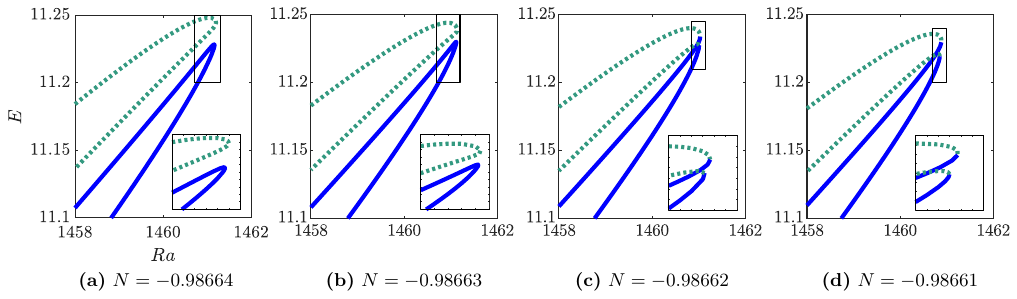}
  \caption[Bifurcation diagrams showing the unfolding around a transcritical bifurcation between $L^-$ and $\tilde{L}^-$ at ${N \approx -0.98662}$]{Bifurcation diagrams showing the unfolding around a transcritical bifurcation between $L^-$ (blue solid) and $\tilde{L}^-$ (green dotted) at ${N \approx -0.98662}$ (near (II) in figure~\ref{fig:pinning_summary}). 
    The branch segments show convectons and multiconvectons with four central rolls close to the right edge of the pinning region.}
  \label{fig:snakestoisola}
\end{figure}
Figures~\ref{fig:bdn0975}(a) and (b) present the bifurcation diagram for $N=-0.975$ after the breakup process has finished and all convectons lie on one of the vertically stacked isolas.
Each isola is `C-shaped' and most are bounded between the left edge of the anticonvecton pinning region at $Ra\approx 1288$ and the first right saddle-node of the anticonvecton branch at $Ra \approx 1314$.
We find that exceptions arise, however, both for the isolas containing convectons with one or two central rolls and those containing nearly domain-filling states, which are found to lie within different ranges of Rayleigh numbers, but note that this may be anticipated as the top and bottom saddle-nodes of regular snaking branches can be located at different values of the parameter from those of the pinning region (see figure~\ref{fig:N099}).

\begin{figure}
  \centering
  \includegraphics[width=0.99\linewidth]{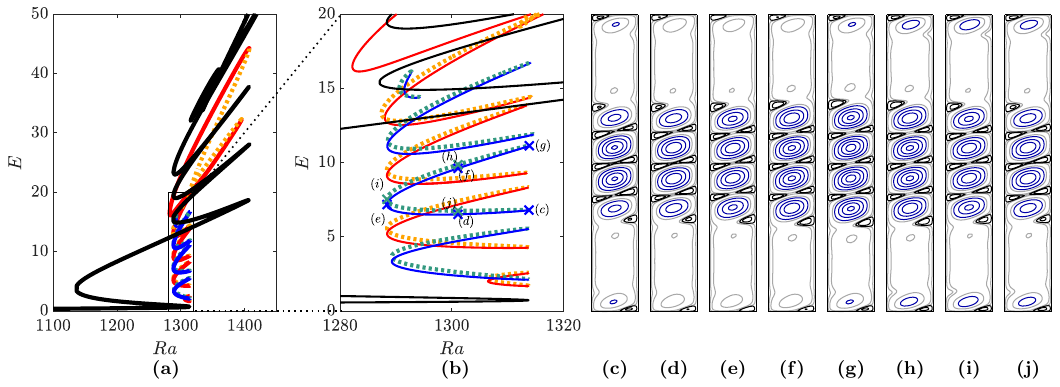}
  \caption[Bifurcation diagram and streamfunctions for ${N = -0.975}$ showing the branch of anticonvectons and stacked isolas of convectons]{(a), (b) Bifurcation diagram for ${N = -0.975}$ showing the branch of anticonvectons (black) and stacked isolas of convectons (red/orange dotted and blue/green dotted).
    (b) Magnification of the vertically stacked isolas. 
    (c)\textendash{}(j) Streamfunctions of selected convectons along the isola with four central rolls.
    Contours are shown using linear (blue) and logarithmic (black and grey) scales and take the values: $[-10^{-3}, -10^{-2}, -10^{-1}]$ (grey); $[10^{-3}, 10^{-2}, 10^{-1}]$ (black); $[-0.2,-0.4,-0.6,...]$ (blue).}
  \label{fig:bdn0975}
\end{figure}

Figures~\ref{fig:bdn0975}(c)\textendash{}(j) illustrate the typical manner in which the form of convectons changes as a single isola is traversed.
Starting at the lower right saddle-node (c) and following the lower branch segment (originally $L^-$, blue) towards the following left saddle-node, the pair of outer central rolls strengthen, while the inner central rolls and end rolls weaken (figures~\ref{fig:bdn0975}(c)\textendash{}(e)).
All six rolls of the convecton proceed to strengthen as this branch segment is followed to the upper right saddle-node (figures~\ref{fig:bdn0975}(e)\textendash{}(g)).
The remaining two segments of the isola (green dotted) originated from the multiconvecton branch, $\tilde{L}^-$ and, as such, we see that the end rolls of the convectons are stronger on these branch segments and have maximal strength at the left saddle-node (i). 
Following these multiconvecton branch segments from right saddle-node (g) back to (c), we see how the central rolls vary in the reverse order to that just described, with the four central rolls first weakening towards the left saddle-node (i).
The outer pair of central rolls continue to weaken from the left saddle-node (i) towards the right saddle-node (c), while the inner central rolls strengthen again.
This behavior echoes that seen in figure~\ref{fig:N099}, where the central rolls within a multiconvecton varied over a single snaking oscillation in the same way as convecton rolls.

\subsubsection{(III): End of convectons}

\begin{figure}
  \centering
  \includegraphics[]{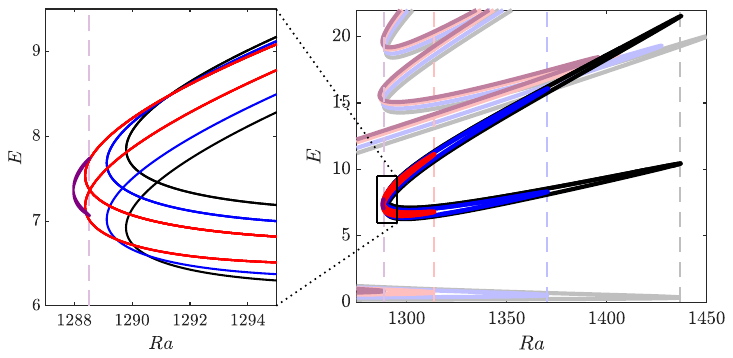}
  \caption[Isolas of convectons with four central rolls as $N$ increases to ${N=-0.9725}$]{Isolas of convectons with four central rolls as $N$ increases to point (III) of figure~\ref{fig:pinning_summary} showing ${N = -0.985}$ (black), ${N=-0.98}$ (blue), ${N=-0.975}$ (red) and ${N = -0.9725}$ (purple).
    The faint solid lines present the anticonvecton branch at the corresponding parameter values, while the vertical dashed lines mark the Rayleigh numbers of the first right saddle-node of each branch.
    The panel on the left shows a magnification of the region near the left saddle-nodes of the isolas.}
  \label{fig:decreasingisolas}
\end{figure}
The first right saddle-node of the anticonvecton branch continues towards lower Rayleigh numbers as the buoyancy ratio increases and ultimately passes through the left edge of the anticonvecton pinning region, as seen in figure~\ref{fig:anticonvectonN}.
The isolas of convectons are bounded by this saddle-node and the left edge of the anticonvecton pinning region, as evidenced in figure~\ref{fig:decreasingisolas}.
Thus, as the buoyancy ratio increases between $N\approx -0.9866$ (II) and $N\approx -0.972$ (III), the right saddle-nodes of the isolas move towards lower Rayleigh numbers.
Meanwhile, the left saddle-nodes only undergo small changes in location, as seen in the left panel of figure~\ref{fig:decreasingisolas}.
The isolas therefore become smaller until they disappear around ${N\approx -0.972}$ (III), when the first right saddle-node of the anticonvecton branch coincides with the left edge of the anticonvecton pinning region.
This behavior explains why we were unable to track the four-roll convecton in figure~\ref{fig:convecton_1350} to $N=-0.97$.

\subsubsection{(IV) and (V): End of anticonvectons}

\begin{figure}
  \centering
  \includegraphics[]{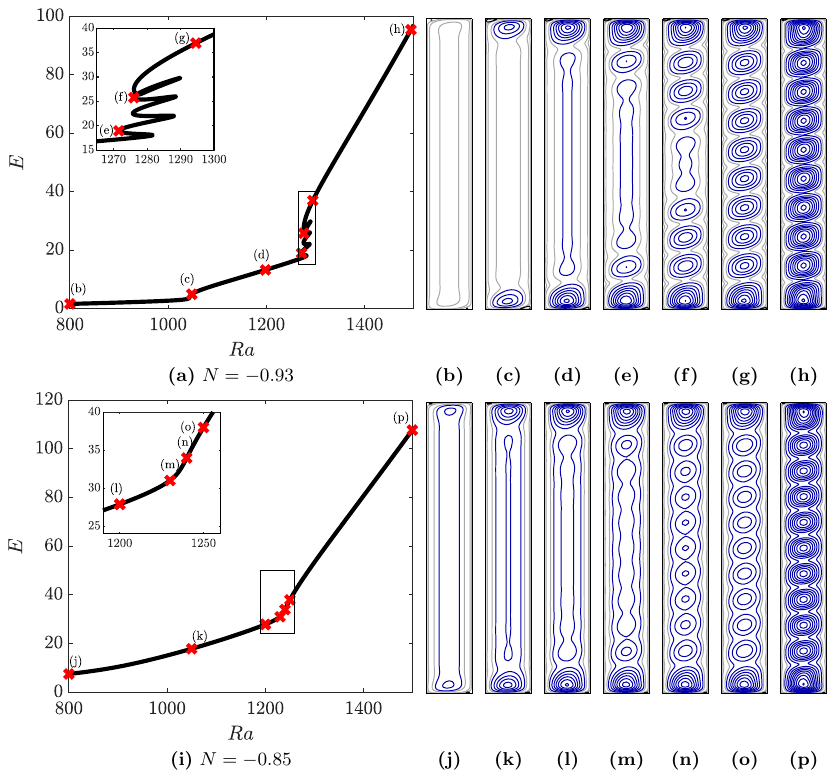}
  \caption[Anticonvecton branches and streamfunctions for selected states along the branch when ${N = -0.93}$ and ${N=-0.85}$]{Anticonvecton branches and streamfunctions for selected states along the branch when ${N = -0.93}$ (top row) and ${N=-0.85}$ (bottom row).
    These represent the behavior either side of point (V) in figure~\ref{fig:pinning_summary}, where the anticonvecton branch snakes for ${N < -0.91}$ and does not snake for ${N > -0.91}$.
    Contours are shown using linear (blue) and logarithmic (black and grey) scales and take the values: $[-10^{-3}, -10^{-2}, -10^{-1}]$ (grey); $[10^{-3}, 10^{-2}, 10^{-1}]$ (black); $[-0.2,-0.4,-0.6,...]$ (blue).}
  \label{fig:anticonvecton_termination}
\end{figure}
The snaking structure of the anticonvecton branch persists over a wider range of buoyancy ratios; its existence, however, still depends upon there being competition between thermal and solutal buoyancy effects and we cease to find snaking anticonvectons when ${N > -0.91}$ (point (V) in figure~\ref{fig:pinning_summary}).

The two-roll anticonvectons (A2) are the first anticonvectons to stop existing.
This occurs by the first right saddle-node of the anticonvecton branch moving towards lower Rayleigh numbers at a faster rate than the following left saddle-node, as seen in figure~\ref{fig:pinning_summary}.
The two saddle-nodes then coincide and disappear in a cusp bifurcation around ${N\approx -0.93}$ (point (IV)).
Figure~\ref{fig:anticonvecton_termination}(a) depicts the anticonvecton branch for $N = -0.93$, shortly after the cusp bifurcation, where the solution labeled (c) is located in the vicinity of this cusp bifurcation and the branch only starts snaking around ${Ra \approx 1281}$.
The branch segment preceding the snaking corresponds to the large-scale anticlockwise flow strengthening and developing stronger secondary rolls at both ends of the domain (figures~\ref{fig:anticonvecton_termination}(b)\textendash{}(d)).
Since these end rolls are secondary rolls of the domain-filling flow, rather than separated from a weaker recirculation flow, these states are dynamically different from the two-roll anticonvectons (A2) illustrated in figure~\ref{fig:pinning_summary}.

The anticonvecton pinning region becomes smaller as the buoyancy ratio increases, as first shown in figures~\ref{fig:anticonvectonN} and \ref{fig:pinning_summary}.
This continues until ${N \approx -0.91}$ (point (V) in figure~\ref{fig:pinning_summary}), when the anticonvecton branch stops snaking.
This transition occurs in the form of successive cusp bifurcations where left and right saddle-nodes disappear.
This process is similar to the way localized states in the quadratic-cubic Swift\textendash{}Hohenberg equation disappear in the approach to the codimension-two point where the criticality of the primary bifurcation changes \cite{burke2006localized}.

Figure~\ref{fig:anticonvecton_termination} illustrates the differences between the evolution of the states along the snaking anticonvecton branch (${N=-0.93}$, top row) and their evolution along the non-snaking branch (${N = -0.85}$, bottom row).
In both cases, convection rolls first appear near the end-walls before new rolls successively appear, filling the domain toward its center.
When ${N = -0.93}$, this nucleation occurs symmetrically pairwise between successive left saddle-nodes (figures~\ref{fig:anticonvecton_termination}(d)\textendash{}(f)).
This continues until a domain-filling ten-roll state is reached (figure~\ref{fig:anticonvecton_termination}(g)), whose rolls proceed to strengthen with increasing Rayleigh number (figure~\ref{fig:anticonvecton_termination}(h)).
However, when ${N=-0.85}$, while we see initial strengthening of rolls adjacent to the end rolls (figure~\ref{fig:anticonvecton_termination}(m)), further interior rolls proceed to strengthen in a spatially modulated manner, rather than in a successive nucleation process, to reach an eleven-roll domain-filling state (figure~\ref{fig:anticonvecton_termination}(n)).
Again, the interior rolls strengthen with Rayleigh number to achieve nearly uniform amplitude by ${Ra = 1500}$ (figures~\ref{fig:anticonvecton_termination}(o) and (p)). 

A weak anticlockwise flow surrounds each of the domain-filling states in figure~\ref{fig:anticonvecton_termination}.
This flow strengthens with increasing $N$, as can be seen by comparing the outer contour lines in figures~\ref{fig:anticonvecton_termination}(h) and (p), while the secondary interior rolls weaken.
While we have not studied these states further into the thermally dominated regime, we would expect the trend to continue until a single domain-filling anticlockwise roll, driven primarily by the thermal buoyancy force, is achieved. 

\section{Discussion}\label{sec:discussion}

In this paper, we have considered how breaking the balance between opposing thermal and solutal gradients affects the formation of patterns in natural doubly diffusive convection within a closed vertical cavity.
In the balanced system, the steady conduction state loses stability via bifurcations which give rise to states comprised of steady convection rolls in the center of the domain known as convectons.
Other types of localized convection states exist which are not connected to the conduction state.
In the absence of balance between thermal and solutal effects, the steady conduction state does not exist and is, instead, replaced at low Rayleigh numbers by a large-scale recirculating flow that strengthens with both increasing Rayleigh number and variation of $N$ away from the balanced system value: ${N = -1}$.
The primary bifurcations of the conduction state, from which branches of convectons bifurcate when ${N = -1}$, unfold as a result of this large-scale flow.
These small-amplitude changes are found to be the main structural differences between the bifurcation diagrams for the balanced and weakly imbalanced systems.
Further differences develop as the buoyancy ratio is increased into the thermally dominated regime as anticlockwise rolls at the ends of the domain tend to strengthen.
This strengthening enables the branch of anticonvectons (steady states comprised of convection rolls located near the end-walls) to connect with the branch of large-scale flow originating from $Ra=0$ by ${N \approx -0.9904}$.
We found that the position of the first right saddle-node of the resulting branch was critical in understanding the evolution of the branches of localized states.
The branches of localized states maintain an organised structure despite variations in the buoyancy ratio.
This differs from related systems with a broken symmetry, including \cite{jacono2017localized,mercader2019effect}, where the structure of the snaking branches become increasingly complex as the degree of symmetry breaking increases.
We attribute this difference to the fact that breaking the balance between thermal and solutal effects preserves the centro-symmetry in this study.

Anticonvectons persist over a wider range of buoyancy ratios than convectons, provided $N > -1$ and we found that they cease to exist as the snaking of their solution branch vanishes.
The anticonvectons, which were characterized by the successive nucleation of interior rolls during snaking, then become states for which all interior rolls strengthen simultaneously as the branch is continued monotonically in Rayleigh number.
This resembles the transition from successive to simultaneous roll formation found in fluids heated laterally within a vertically stabilizing gradient of salt, where Lee et al. \cite{lee1990confined} used experimental data to divide the parameter space consisting in thermal and solutal Rayleigh numbers into regimes corresponding to these two types of behavior together with stagnant and unicellular flows.
They found that the boundaries between regimes occurred at approximately constant buoyancy ratios between ${N\approx 10}$ and ${N\approx 55}$, which Lee and Hyun \cite{lee1991double} later confirmed numerically and Dijkstra and Kranenborg \cite{dijkstra1996bifurcation} related to paths of bifurcation points.
We should note, however, that these studies considered different doubly diffusive convection setups (with $N>0$) and, hence, we should not try to directly compare results.


While we have identified how the branches of localized states terminate in the thermally dominated regime of doubly diffusive convection, our results indicate that they can be expected to exist on a broader parameter range in the solutally dominated regime (Figure \ref{fig:pinning_summary}).
In this regime, the large-scale recirculating flow now opposes the anticlockwise flow of the central convecton rolls, which will lead to more complex dynamics.
To begin to explore the potential differences, we tracked a four-roll convecton at ${Ra = 1350}$ from ${N = -1}$ into both the thermally and solutally dominated regimes and present the streamfunctions of these states from ${N = -0.98}$ to ${N =-1.2}$ in figure~\ref{fig:convecton_1350}.
\begin{figure}
  \centering
  \includegraphics[width=\linewidth]{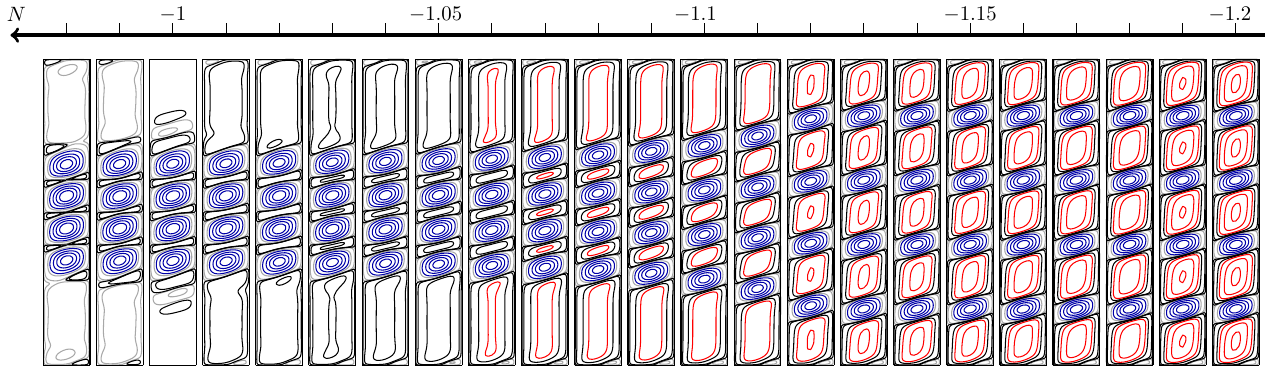}
  \caption[Convectons at ${Ra = 1350}$ and ${Pr = 1}$, with the buoyancy ratio varying in intervals of ${N=-0.01}$ from ${N = -0.98}$ to ${N = -1.2}$]{Convectons at ${Ra = 1350}$ and ${Pr = 1}$, with the buoyancy ratio varying in intervals of ${N=-0.01}$ from ${N = -0.98}$ to ${N = -1.2}$.
    Contours are shown using linear (blue and red) and logarithmic (black and grey) scales and take the values: $[-10^{-3}, -10^{-2}, -10^{-1}]$ (grey); $[10^{-3}, 10^{-2}, 10^{-1}]$ (black); $[-0.2,-0.4,-0.6,...]$ (blue); $[0.2, 0.4, 0.6,...]$ (red).}
  \label{fig:convecton_1350}
\end{figure}
Upon increasing $N$ into the thermally dominated regime, the central convecton rolls undergo similar changes to when the Rayleigh number was increased in the balanced (figure~\ref{fig:convecton_branch}) and weakly imbalanced (figure~\ref{fig:N0999}) regimes.
The small clockwise rolls that separate the four strong anticlockwise rolls also weaken and split, which leads to the state ceasing to exist by ${N = -0.97}$.
Upon decreasing $N$ into the solutally dominated regime from $N=-1$, instead of the convectons ceasing to exist, they evolve continuously into domain-filling states consisting of counter-rotating rolls through the strengthening of the clockwise rolls, as seen in figure~\ref{fig:convecton_1350}.
A similar transition to different domain-filling patterned states may also be observed starting from convectons with different numbers of central rolls.
This opens new fundamentally important questions related to the potential variety of domain-filling states and the transition to them from localized states, which will be addressed in future work.

The results presented in this paper focused on a domain size of $12$ critical wavelengths, but we anticipate that the phenomena unraveled on convectons, anticonvectons and multiconvectons, including the unfolding of their bifurcations, extends to any larger domain size provided the localized states are not close to be domain-filling.

\begin{acknowledgments}
This work was undertaken on ARC3 and ARC4, part of the High Performance Computing facilities at the University of Leeds, UK.
The first author is grateful for support from the Leeds-York Natural Environment Research Council
(NERC) Doctoral Training Partnership (DTP) SPHERES under grant NE/L002574/1.
For the purpose of open access, the authors have applied a Creative Commons Attribution (CC~BY) licence to 
any Author Accepted Manuscript version arising from this submission.
\end{acknowledgments}


%

\end{document}